%% file: ms.tex
\documentclass[apj]{emulateapj}
\usepackage{apjfonts}
\usepackage{epsfig}
\usepackage{amsmath}
\usepackage{mathrsfs}

\newcommand{\be}{\begin{eqnarray}}
\newcommand{\ee}{\end{eqnarray}}
\newcommand{\beq}{\begin{equation}}
\newcommand{\eeq}{\end{equation}}
\def\simless{\mathbin{\lower 3pt\hbox
      {$\rlap{\raise 5pt\hbox{$\char'074$}}\mathchar"7218$}}}
\def\simgreat{\mathbin{\lower 3pt\hbox
      {$\rlap{\raise 5pt\hbox{$\char'076$}}\mathchar"7218$}}} %> or of order

\renewcommand{\vec}[1]{\mathbf{#1}}
\newcommand{\xivec}{{\mbox{\boldmath $\xi$}}}

\newcommand{\grad}{\boldsymbol{\nabla}}

\newcommand{\pc}{\,{\rm pc}}

\newcommand{\msun}{\,M_\odot}
\newcommand{\rsun}{\,R_\odot}

\newcommand{\fdisk}{\mathscr{F}}
\newcommand{\pert}{\varepsilon}

\begin{document}

% -----------------------------------------------------------
% -----------------------------------------------------------

\shorttitle{ELLIPSOIDAL OSCILLATIONS}
\shortauthors{PFAHL, ARRAS, \& PAXTON}

% -----------------------------------------------------------
% -----------------------------------------------------------

\title{Ellipsoidal Oscillations Induced by Substellar Companions: \\
A Prospect for the {\em Kepler} Mission}

\author{Eric Pfahl\altaffilmark{1}, Phil Arras\altaffilmark{1,2}, 
  and Bill Paxton\altaffilmark{1}}

\altaffiltext{1}{Kavli Institute for Theoretical Physics, University
of California, Santa Barbara, CA 93106; pfahl@kitp.ucsb.edu,
paxton@kitp.ucsb.edu}

\altaffiltext{2}{Department of Astronomy, University of Virginia,
P.O. Box 400325, Charlottesville, VA 22904-4325; arras@virginia.edu}

% -----------------------------------------------------------
% -----------------------------------------------------------

\begin{abstract}

  Hundreds of substellar companions to solar-type stars will be
  discovered with the {\em Kepler} satellite.  {\em Kepler}'s extreme
  photometric precision gives access to low-amplitude stellar
  variability contributed by a variety of physical processes.  We
  discuss in detail the periodic flux modulations arising from the
  tidal force on the star due to a substellar companion.  An analytic
  expression for the variability is derived in the equilibrium-tide
  approximation.  We demonstrate analytically and through numerical
  solutions of the linear, nonadiabatic stellar oscillation equations
  that the equilibrium-tide formula works extremely well for stars of
  mass $<$$1.4\msun$ with thick surface convection zones.  More
  massive stars with largely radiative envelopes do not conform to the
  equilibrium-tide approximation and can exhibit flux variations
  $\ga$10 times larger than naive estimates.  Over the full range of
  stellar masses considered, we treat the oscillatory response of the
  convection zone by adapting a prescription that A. J. Brickhill
  developed for pulsating white dwarfs.  Compared to other sources of
  periodic variability, the ellipsoidal lightcurve has a distinct
  dependence on time and system parameters.  We suggest that
  ellipsoidal oscillations induced by giant planets may be detectable
  from as many as $\sim$100 of the $10^5$ {\em Kepler} target stars.
  For the subset of these stars that show transits and have
  radial-velocity measurements, all system parameters are well
  constrained, and measurement of ellipsoidal variation provides a
  consistency check, as well as a test of the theory of forced stellar
  oscillations in a challenging regime.
    
\end{abstract}


% -----------------------------------------------------------
% -----------------------------------------------------------

\section{INTRODUCTION}
\label{sec:intro}

The upcoming {\em Kepler}\footnote{\url{http://kepler.nasa.gov}}
satellite will continuously monitor $\sim$$10^5$ main-sequence stars
of mass $\simeq$0.5--$1.5\msun$ over 4--6 years with fractional
photometric precisions of $\sim$$10^{-5}$.  Such high sensitivity,
which is unattainable from the ground, will allow for the robust
detection of Earth-size planets that transit their host stars, and the
measurement of asteroseismic oscillations as a probe of stellar
structure \citep[e.g.,][]{Borucki2004,Basri2005}.  These missions will
also discover hundreds of ``hot Jupiters'' with orbital periods of
$<$10 days, revealed by their transits or reflected starlight
\citep[e.g.,][]{Jenkins2003}.  Continuous observations of these
systems are likely to show a myriad of novel physical effects,
including Doppler flux variability of the host stars \citep{Loeb2003},
photometric dips due to moons or rings around the planets
\citep{Sartoretti1999,Brown2001}, and the impact of additional
perturbing planets on transit timing
\citep{Miralda-Escude2002,Agol2005,Holman2005}.  The same ideas apply
if the companion is a more massive brown dwarf, but these are rarely
found in close orbits around solar-type stars
\citep[e.g.,][]{Grether2006}.

Here we scrutinize another mechanism for generating periodic
variability of a star closely orbited by a giant planet or brown
dwarf. A star subject to the tidal gravity of a binary companion has a
nonspherical shape and surface-brightness distribution.  In the
simplest approximation, the stellar surface is a prolate ellipsoid
with its long axis on the line connecting the two objects.  As the
tidal bulge tracks the orbital motion, differing amounts of light
reach the observer.  For a solar-type star orbited by a perturbing
companion of mass $M_p$ with period $P_{\rm orb}$, the expected
fractional amplitude of this ellipsoidal variability is
$\sim$$10^{-2}(M_p/M_\odot)(1\,{\rm day}/P_{\rm orb})^2$.  This effect
has a long history in the study of eclipsing binary stars \citep[see
the review by][]{Wilson1994}, but was mentioned only recently in the
exoplanet context.

\citet{Udalski2002}, \citet{Drake2003}, and \citet{Sirko2003} noted
that if ellipsoidal light variations are detected from the ground,
where the fractional photometric precision is $\ga$$10^{-3}$, then the
perturber must be fairly massive (e.g., $\ga$$0.1\msun$).  They
offered this idea as a test to distinguish between planetary transits
and eclipses by low-mass stars.  The superior sensitivity of {\em
  Kepler} offers the possibility of measuring ellipsoidal variability
induced by giant planets ($M_p \sim 10^{-3}$--$10^{-2}\msun$) with 
orbital periods of $\la$10\,days. 

\citet{Loeb2003} compare the ellipsoidal variability induced by a
planetary companion to flux modulations arising from reflected
starlight and the Doppler effect. The three amplitudes are similar
when the companion has an orbital period of $\la$3 days and an optical
albedo of $\la$0.1.  In a sufficiently long observation it should be
possible to separately extract each of the signals, since their
Fourier decompositions are distinct.  Precise physical modeling of the
ellipsoidal lightcurve could provide an independent constraint on the
mass of the companion, as well as important clues regarding stellar
tidal interactions.

Ellipsoidal variability is typically modeled under the assumption that
the distorted star maintains hydrostatic balance and precisely fills a
level surface of an appropriate potential (e.g., the Roche potential).
The measured flux is then just an integral of the intensity over the
visible stellar surface, where the intensity includes the effects of
limb darkening and gravity darkening \citep[e.g.,][]{Kopal1942}.  This
approach is strictly valid only when the orbit is circular and the
star rotates at the orbital frequency, so that a stationary
configuration exists in the coorbital frame.  These conditions may not
be satisfied when the companion has a low mass or long period, because
of the weak tidal interaction.  In fact, a state of tidal equilibrium
may not be attainable in the case of a planetary companion
\citep[e.g.,][]{Rasio1996}.  Equilibrium models of ellipsoidal
lightcurves do have a realm of validity for noncircular orbits and
asynchronously rototating stars, and have been applied successfully to
somewhat eccentric binaries \citep[e.g.,][]{Soszynski2004}.  However,
by construction, such models ignore fluid inertia and the possibility
exciting normal modes of oscillation, effects that may be of critical
importance in a wide range of observationally relevant circumstances.
Here we apply the machinery of linear stellar oscillation theory to
the weak tidal forcing of stars by substellar companions.
Conceptually, our investigation bridges {\em Kepler}'s planetary and
astroseismology programs.

Section~2 describes the geometry of the problem, provides quantitative
measures for the strength of the tidal interaction, discusses our
simplifying assumptions, and presents the mathematical framework for
calculating ellipsoidal variability. In \S~\ref{sec:eqtide}, we
consider the equilibrium-tide approximation and derive an analytic
expression for the ellipsoidal lightcurve.  A brief review of von
Zeipel's theorem and its limitations is given in
\S~\ref{sec:vonzeipel}.  Tidally forced, nonadiabatic stellar
oscillations are addressed in \S~\ref{sec:nonad}, where we argue for a
simple treatment of perturbed surface convection zones, use this
prescription to calculate the ellipsoidal variability of deeply
convective stars, estimate analytically the surface flux perturbation
in mainly radiative stars, and show select numerical results.  Our
main conclusions are summarized in \S~\ref{sec:summary}.  We conclude
in \S~\ref{sec:detection} with remarks on the measurement of
ellipsoidal oscillations in the presence of other sources of periodic
variability.

% -----------------------------------------------------------

\section{PRELIMINARIES}
\label{sec:prelim}

Consider a star of mass $M$ and radius $R$ is orbited by a substellar
companion of mass $M_p$ and radius $R_p$.  We work in spherical
coordinates $(r, \theta, \phi)$ with the origin at the star's center
and the pole direction ($\theta = 0$) parallel to the orbital angular
momentum vector.  The orbit is then described by $(d,
\pi/2, \phi_p)$, where $d$ and $\phi_p$ are, respectively, the
time-dependent orbital separation and true anomaly; $\phi_p = 0$ marks
the phase of periastron.  We assume that the orbit is strictly
Keplerian with fixed semimajor axis $a$ and eccentricity $e$, such
that $d = a(1 - e^2)/(1 + e\cos \phi_p)$.  The direction to the
observer from the center of the star is $(\theta_o, \phi_o)$, so that
the conventional orbital inclination is $I = \pi - \theta_o$.

We imagine that the gravity of the companion raises nearly symmetrical
tidal bulges on opposite sides of the star that rotate at the orbital
frequency.  A rough measure of both the height of the tides relative
to the unperturbed stellar radius and the fractional amplitude of the
ellipsoidal variability is given by the ratio of the tidal
acceleration to the star's surface gravity:
\be\label{eq:tidescale}
\pert \equiv \frac{M_p}{M}\left( \frac{R}{a} \right)^3 \sim
10^{-5}\, \frac{M_p}{M_J} \frac{M_\odot}{M}
\left( \frac{P_*}{2.8\ {\rm hr}}\frac{1\ {\rm day}}{P_{\rm orb}} \right)^2 
~,
\ee
where $M_J \simeq 10^{-3}\msun$ is the mass of Jupiter,
$P_*=2\pi(R^3/GM)^{1/2}=2.8\,[(R/R_\odot)^{3}(M_\odot/M)]^{1/2}\,{\rm
  hr}$ is the dynamical time of the star.  For main-sequence stars
with $R/R_\odot \simeq M/M_\odot$, we see that $\pert \propto M_p M
P_{\rm orb}^{-2}$.  The maximum value of $\pert$ is attained when the
companion fills its Roche lobe at an orbital separation of $a \simeq 2
R_p(M/M_p)^{1/3}$, which gives
\be
\pert_{\rm max} &\simeq&
\left(\frac{M_p}{M}\right)^2
\left(\frac{R}{2R_p}\right)^3 \nonumber \\
&\simeq&
10^{-4}\,\left(\frac{M_p}{M_J}\right)^2
\frac{M}{M_\odot}
\left(\frac{0.1\rsun}{R_p}\right)^3
~,
\ee
where we have applied a fixed value of $R_p = 0.1\rsun$, appropriate
for both giant planets and old brown dwarfs.  Note that $\pert_{\rm
  max}\sim 1$ for massive brown dwarfs ($M_p/M_J \sim 80$).
Hereafter, we consider only cases with $\pert \ll 1$.

For orbital periods as short as $\simeq$1 day, tidal torques on the
star from a planetary companion are rather ineffective at altering the
stellar rotation rate \citep[e.g.,][]{Rasio1996}.  Therefore, as
already mentioned in \S~\ref{sec:intro}, we should not generally
expect the star to rotate synchronously with the orbit, and so there
is no frame in which the star appears static.  This holds when the
orbit is circular, and is obviously true when the there is a finite
eccentricity.  In fact, $\simeq$30\% of the known
exoplanets\footnote{\url{http://vo.obspm.fr/exoplanetes/encyclo/encycl.html}}
with $P_{\rm orb} < 10$\,days have eccentricities of $>$0.1. Small
variable distortions of the star from its equilibrium state, due to a
combination of asynchronous rotation and orbital eccentricity, should
be viewed as waves excited by the tidal force of the companion.  Our
task is to study such tidally forced stellar oscillations in the
linear domain in order to understand the corresponding lightcurves.

In order to greatly simplify the mathematical description of the
stellar oscillations, we assume that the star is nonrotating in the
inertial frame.  When the stellar rotation frequency is nonzero, but
much smaller than the tidal forcing frequency, the effect of rotation
is to introduce fine structure into the oscillation frequency
spectrum, and cause the oscillation eigenfunctions to be slightly
modified as a result of the Coriolis force \citep[for a 
discussion, see][]{Unno1989}.  Tidal pumping of a slowly rotating star
by an orbiting companion has a dominant period of $P_{\rm orb}/2$---a
few days in the cases of interest. By contrast, single solar-type
stars with ages $>$1\,Gyr tend to have rotation periods of $>$10 days
\citep[e.g.,][]{Skumanich1972,Pace2004}; the Sun has an equatorial
rotation period of $\simeq$25 days.  Slowly rotating stars with masses
of $\simeq$$1\msun$ are prime targets for {\em Kepler}, since they
exhibit low intrinsic variability. Based on this selection effect, and
the inability of tidal torques to spin up the star, our assumption of
vanishing stellar rotation seems generally justified.

The general framework for calculating the measurable flux modulations
associated with ellipsoidal stellar oscillations is as follows.  We
consider small perturbations to a spherical, nonrotating background
stellar model, such that fluid elements at equilibrium position
$\vec{x}$ are displaced in a Lagrangian fashion to position $\vec{x} +
\xivec$.  Variations in the measured flux from an oscillating star
arise from two physically distinct contributions
\citep[e.g.,][]{Dziembowski1977}: (1) changes in the shape of the star
due to radial fluid displacements $\xi_r=\xivec\cdot \vec{e}_r$, where
$\vec{e}_r$ is the radial unit vector, and (2) hot and cold spots
generated by local Lagrangian perturbations $\Delta F$ to the heat
flux.  Our main task in \S\S~\ref{sec:eqtide} and \ref{sec:nonad} is
to compute $\xi_r$ and $\Delta F$ according to the relevant physics.

Given the dependences of $\xi_r$ and $\Delta F$ on $(r,\theta,\phi)$,
it is straightforward to compute the time varying component of the
measured flux.  The flux\footnote{Our calculations concern the
  bolometric flux, although is relatively straightforward to modify
  the analysis for narrow-band measurements.} received from a star at
distance $D$ is \citep[e.g.,][]{Robinson1982}
\be
\fdisk & = &  
\frac{1}{D^2} \int dS\ 
\vec{n} \cdot \vec{n}_o \ F \ h(\vec{n} \cdot \vec{n}_o)
\ee
where $dS$ is an area element at the stellar photosphere, $F$ is the
net flux of radiation out of the surface element, $h$ is the
limb-darkening function, $\vec{n}$ and $\vec{n}_o$ are unit vectors
normal to the surface and toward the observer, respectively, and the
integration is over the visible stellar disk.  Vertical displacement
at the surface yields changes in $\fdisk$ through changes in surface
area and $\vec{n}\cdot \vec{n}_o$.  Following \citet{Dziembowski1977},
we expand $\xi_r$ and $\Delta F$ in spherical harmonics,
\be
\xi_r(r,\theta,\phi,t) & = & \sum_{\ell = 0}^\infty \sum_{m = -\ell}^\ell 
\xi_{r, \ell m}(r,t) Y_{\ell m}(\theta,\phi)~, \\
\Delta F(r,\theta,\phi,t) & = & \sum_{\ell = 0}^\infty \sum_{m = -\ell}^\ell
\Delta F_{\ell m}(r,t) Y_{\ell m}(\theta,\phi)~,
\ee
and carry out the appropriate linear expansions to obtain the
fractional variability
\be\label{eq:fdiskgen}
\frac{\delta\fdisk}{\fdisk} =
\sum_{\ell=0}^\infty
\left[ (2b_\ell - c_\ell) \frac{\xi_{r,\ell}^o}{R} + 
b_\ell \frac{\Delta F_\ell^o}{F}  \right]~.
\label{eq:dflux}
\ee
Here $\xi_{r, \ell}^o$ and $\Delta F_\ell^o$ are components evaluated
at the surface ($r = R$) and in the direction of the observer:
\be\label{eq:xirdelf_obs}
\frac{\xi_{r, \ell}^o}{R} & = &
\sum_{m = -\ell}^\ell \frac{\xi_{r, \ell m}(R,t)}{R} 
Y_{\ell m}(\theta_o,\phi_o)~, \\
\frac{\Delta F_\ell^o}{F} & = & 
\sum_{m = -\ell}^\ell \frac{\Delta F_{\ell m}(R,t)}{F} 
Y_{\ell m}(\theta_o,\phi_o)~.
\ee
The terms $b_\ell$ and $c_\ell$ are given by
\be
b_\ell =
\int_0^1 d\mu \mu\, P_\ell\, h~,~~
c_\ell = 
\int_0^1 d\mu (1-\mu^2)\frac{dP_\ell}{d\mu}
\left( h + \mu \frac{dh}{d\mu} \right)\ ,
\ee
where $\mu = \vec{n}\cdot\vec{n}_o$, the $P_\ell(\mu)$ are ordinary
Legendre polynomials, and $h(\mu)$ is normalized such that $\int_0^1
d\mu \mu\,h = 1$.  The linear limb-darkening function is
\be\label{eq:limbdark}
h(\mu) = \frac{6}{(3-\gamma)}\big[ 1 - \gamma(1 - \mu)\big] ~;
\ee
more general nonlinear functions of $\mu$ \citep[e.g.,][]{Claret2000}
will not be considered here.  The classical Eddington limb-darkening
function is $h = 1 + 3\mu/2$ \citep[$\gamma = 3/5$;
e.g.,][]{Mihalas1970}.  Table~\ref{tab:ldterms} shows shows functional
forms and particular values of $b_\ell$ and $c_\ell$ for $\ell = 2$
and 3.

% -----------------------------------------------------------

\section{EQUILIBRIUM TIDE}
\label{sec:eqtide}

Vertical displacement of the stellar surface is often accurately
modeled by assuming that the tidally perturbed fluid remains in
hydrostatic balance. The cause and magnitude of the surface flux
perturbation is a more complicated affair.  In this section, we apply
a simple parameterization of the flux perturbation and obtain a
complete set of formulae for computing the ellipsoidal lightcurve.
Subsequent sections provide more detailed calculations.  In
particular, we show in \S~\ref{sec:thickconv} that stars with deep
convective envelopes (the majority of {\em Kepler} targets) have
surface flux variations that conform to the equilibrium-tide
approximation.

\input{tab1.tex}

When the tidal forces on the stellar fluid change sufficiently slowly,
the star can stay very nearly in hydrostatic equilibrium.  If the net
acceleration required to balance the pressure gradient is derivable
from a potential, then equilibrium implies that a fluid element
remains on an equipotential surface.  Since we neglect stellar
rotation, there is no centrifugal force, and the total potential is
the sum of the gravitational potential $\varphi$ from the spherical
background stellar model and the perturbing tidal potential $U \sim
\pert\varphi \ll \varphi$.  For our analytic work, we neglect the
modification of $\varphi$ due to the tide.  In general, the Eulerian
variation $\delta\varphi$ should be added to $U$, as we do in our
numerical models (see \S~\ref{sec:numerical} and the Appendix); we
find that $|\delta\varphi|/|U| \sim 10^{-2}$.

In the absence of tidal forces, a given fluid element sits at
equilibrium position $\vec{x}$ with total potential
$\varphi(\vec{x})$.  Gentle inclusion of the tidal potential causes
the fluid element to move to position $\vec{x} + \xivec$ while
preserving the value of the total potential. This is expressed
mathematically by
\be
\varphi({\vec{x}})  & = & 
\varphi(\vec{x} + \xivec) + U(\vec{x} + \xivec,t) \nonumber \\
& = & \varphi(\vec{x}) + \xivec\cdot\grad\varphi + U(\vec{x},t) +
{\cal O}(\xi^2, \xi U)~.  \ee
We see that $\xivec\cdot\grad\varphi = \xi_r g$, where $g = GM_r/r^2$
is the background gravitational acceleration at mass coordinate $M_r$.
To first order, the radial displacement of the equilibrium tide is
\citep[see also][]{Goldreich1989}
\be\label{eq:eqtide}
\xi_r(\vec{x},t) & \simeq & - U(\vec{x},t)/g~,
\ee
which tells us the geometry of the star as a function of time.

The tidal potential within the star can be expanded as 
\be\label{eq:U}
U(r,\theta,\phi,t) = - \frac{GM_p}{d}
\sum_{\ell = 2}^\infty 
\left(\frac{r}{d}\right)^{\ell} P_\ell(\cos\psi)~,
\ee
where $\cos\psi = \sin\theta \cos(\phi_p-\phi)$.  There is no $\ell =
1$ term, since this would give the acceleration of the star's center
of mass, which is already incorporated into the orbital dynamics.  The
angular expansion of $\xi_r$ follows immediately from
eq.~(\ref{eq:eqtide}):
\be\label{eq:xireqtide}
\frac{\xi_r(r,\theta,\phi,t)}{r} = \frac{M_p}{M_r}
\sum_{\ell = 2}^\infty 
\left(\frac{r}{d}\right)^{\ell+1} P_\ell(\cos\psi)~.
\ee
In order to express $U$ and $\xi_r$ in spherical harmonics, we utilize
the addition theorem,
\be\label{eq:legendre}
P_\ell(\cos\psi) = 
\frac{4\pi}{2\ell+1}
\sum_{m = -\ell}^\ell 
Y_{\ell m}^*(\pi/2,\phi_p)\
Y_{\ell m}(\theta,\phi)~,
\ee
where ``$*$'' denotes the complex conjugate. Note that $Y_{\ell
  m}(\pi/2,\phi_p)$ is nonzero only when $\ell -m$ is even.  For the
dominant $\ell = 2$ components of $U$ and $\xi_r$, the surface values
of $U/\varphi$ and $\xi_r/R$ are $\sim$$\pert$, as expected.

From eqs.~(\ref{eq:xirdelf_obs}), (\ref{eq:xireqtide}), and
(\ref{eq:legendre}), the components $\xi_{r,\ell}^o/R$ of the surface
radial displacement toward the observer are immediately apparent.  As
we will see in \S~\ref{sec:nonad}, the computation of $\Delta F/F$ is,
in general, rather technical.  However, in the special case where the
stellar fluid responds adiabatically to a slowly varying tidal
potential, $\Delta F_\ell/F$ varies in phase with and in proportion to
$\xi_{r,\ell}/r$ in the linear approximation of the equilibrium tide.
Making this assumption, we write $\Delta F_\ell/F= -\lambda_\ell
\xi_{r,\ell}/R$ at the surface, where the $\lambda_\ell$ are real
constants that depend on the stellar structure (see
\S~\ref{sec:conv}).  We will see in \S~\ref{sec:vonzeipel} that
$\lambda_\ell = \ell + 2$ is a good first guess for radiative stars,
and so we might generally expect $\lambda_\ell$ to be positive and
${\cal O}(\ell)$.

We now have the ingredients for the fractional variability
(eq.~[\ref{eq:fdiskgen}]), and we obtain
\be\label{eq:fdiskeq}
\frac{\delta \fdisk}{\fdisk} = 
\pert \sum_{\ell = 2}^\infty
\Bigg(\frac{R}{a}\Bigg)^{\ell - 2}
\Bigg(\frac{a}{d}\Bigg)^{\ell + 1}
f_\ell \ P_\ell (\cos\psi_o)~,
\ee
where $f_\ell = (2-\lambda_\ell)b_\ell - c_\ell$, and $\cos\psi_o =
\sin\theta_o\cos(\phi_p - \phi_o)$.  The $\ell = 2$ and 3 Legendre
polynomials can be expanded as
\begin{multline}\label{eq:P2}
P_2(\cos\psi_o) = 
\frac{1}{4}
\big[
-(3\cos^2 I - 1) \\
+ 3\sin^2 I\ \cos 2(\phi_p - \phi_o)
\big]~,
\end{multline}
\begin{multline}\label{eq:P3}
P_3(\cos\psi_o) = 
\frac{1}{8}\sin I
\big[
-3(5\cos^2 I - 1)\cos (\phi_p - \phi_o) \\
+ 5\sin^2 I\ \cos 3(\phi_p - \phi_o)
\big]~,
\end{multline}
where we have substituted $\theta_o = \pi - I$.  The Eddington
limb-darkening formula gives (see Table~\ref{tab:ldterms})
\be
f_2 = -\frac{13}{10}\left(1 + \frac{\lambda_2}{4}\right)~,~~
f_3 = -\frac{5}{8}\left(1 + \frac{\lambda_3}{10}\right)~.
\ee
It is important to note that $f_2<0$ when $\lambda_2 \geq 0$ (see
below).

\begin{figure}
\centerline{\epsfig{file = 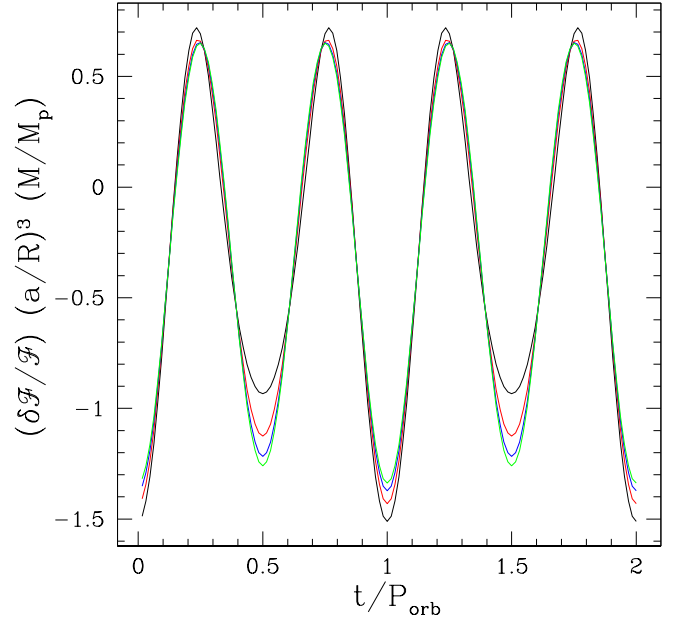,angle = 0,width = \linewidth}}
\caption{ Disk-averaged flux variation (eq.~[\ref{eq:fdiskeq}]) for an
  edge-on circular orbit under the equilibrium-tide approximation
  (eqs.~[\ref{eq:dflux}] and [\ref{eq:xireqtide}]) with $\Delta F/F =
  0$ at the surface. The four curves correspond to $a/R=2$ (black),
  $4$ (red), $8$ (blue) and $16$ (green). In order to compare the
  shapes of the curves, $\delta\fdisk/\fdisk$ has been multiplied by
  $(a/R)^3(M/M_p)$.  As $a/R$ increases, higher harmonics decrease in
  strength and the lightcurve approaches a pure cosine with frequency
  $2/P_{\rm orb}$.  The tidal bulge closest to the companion points
  toward the observer at integer values of $t/P_{\rm orb}$.}
  \label{fig:eqtide_shape} 
\end{figure}

In eq.~(\ref{eq:fdiskeq}), the orbital dynamics are described by the
evolution of $d$ and $\phi_p$ (see \S~\ref{sec:prelim}).  For a
circular orbit, we have $d = a$ and $\phi_p = \Omega t$, where $\Omega
= 2\pi/P_{\rm orb}$, and $t$ is the time since periastron (modulo
$P_{\rm orb}$).  Example lightcurves with $e = 0$, $\gamma = 3/5$,
$\lambda_\ell=0$, and $I = \pi/2$ are shown in
Fig.~\ref{fig:eqtide_shape} for $a/R = \{2, 4, 8, 16\}$.  When $R/a
\ll 1$, the $\ell = 2$ piece of $\delta\fdisk/\fdisk$ is a good
approximation, and the temporal flux variation approaches a pure
cosine with angular frequency $2\Omega$ (see eq.~[\ref{eq:P2}]).
Because $f_2 < 0$, the dominant $\ell = 2$ component of the
ellipsoidal variability has {\em minimum} light when tidal bulge is
aligned with the direction to the observer.  As $R/a$ increases, so
does the importance of $\ell >2$ terms and their extra harmonic
content, as seen in eq.~(\ref{eq:P3}) and Fig.~\ref{fig:eqtide_shape}.

Additional harmonics in $\delta\fdisk/\fdisk$ also result from a
finite eccentricity.  At the ${\cal O}(e)$ level, signals with
frequencies $\Omega$ and $3\Omega$, and amplitudes of $\sim$$\pert e$,
are present in the $\ell = 2$ component of $\delta\fdisk/\fdisk$,
which compete with the $\ell = 3$ piece when $e \sim R/a$.  Notice
that when $e > 0$ the flux is variable even when the orbit is viewed
face-on ($I = 0$ or $\pi$), by virtue of changes in $d^{-3} = 1+3
e\cos\Omega t + {\cal O}(e^2)$.  For $I = 0$, we see that
$P_3(\cos\psi_o)$ vanishes, leaving the largest contribution
$\delta\fdisk/\fdisk \simeq -1.5 \pert e f_2\cos(\Omega t)$.

%

% -----------------------------------------------------------

\section{An Aside on von Zeipel's Theorem}
\label{sec:vonzeipel}

Our equilibrium calculation in the last section used the simple
prescription $\Delta F_\ell/F = - \lambda_\ell \xi_{r,\ell}/R$.  There
remains the question of what physics determines $\Delta F/F$.  A
common practice in empirical studies of close eclipsing
binaries---systems that tend to be nearly in tidal equilibrium---is to
use some variant of the \citet{vonZeipel1924} theorem, which was
originally formulated for purely radiative, strictly hydrostatic
stars.  In equilibrium, all the thermodynamic variables depend only on
the local value of the total potential $\Phi$.  Thus, the radiative
flux can be written as \citep[e.g.,][]{Hansen1994}
\be\label{eq:vonzeipel}
\vec{F} & \propto & - \frac{1}{\kappa \rho} \frac{dT^4}{d\Phi}
\grad \Phi~,
\ee
where $\rho$ is the mass density, $T$ is the effective temperature,
and $\kappa(\rho, T)$ is the opacity.  Equation~(\ref{eq:vonzeipel})
is the essence of von Zeipel's theorem, which says that the magnitude
$F$ of the radiative flux is proportional to the magnitude of the net
acceleration $A = |\grad\Phi|$.  When $\Phi = \varphi + U$ (see
\S~\ref{sec:eqtide}), we obtain  $A = g + \partial U/\partial r + 
{\cal O}(\xi^2)$, so that the Lagrangian flux
perturbation about equilibrium is
\be\label{eq:fluxvar}
\frac{\Delta F}{F} = \frac{\Delta A}{g} = 
\frac{\Delta g}{g} + \frac{1}{g}\frac{\partial U}{\partial r}~,
\ee
where $\Delta g/g = -2\xi_r/r$, due to the change in radius at
approximately constant enclosed mass.  Substituting the
equilibrium-tide result $U = -\xi_r g$ into eq.~(\ref{eq:fluxvar}), we
obtain the compact expression $\Delta F/F = -\partial \xi_r/\partial
r$.  Using eq.~(\ref{eq:xireqtide}), we find
\be\label{eq:fluxvar2}
\frac{\Delta F_\ell}{F} = -(\ell + 2) \frac{\xi_{r,\ell}}{r}~,
\ee
from which we identify $\lambda_\ell = \ell + 2$. 

Although the application of von Zeipel's theorem is instructive, the
underlying physical assumptions are inaccurate for slowly rotating 
main-sequence stars of mass 1.0--$1.6\msun$ with tidal forcing 
periods of days.  We are now led to investigate the general problem 
of forced nonadiabatic stellar oscillations.

% -----------------------------------------------------------

\section{FORCED NONADIABATIC OSCILLATIONS}
\label{sec:nonad}

% -----------------------------------------------------------

%
\begin{figure}
\centerline{\epsfig{file = 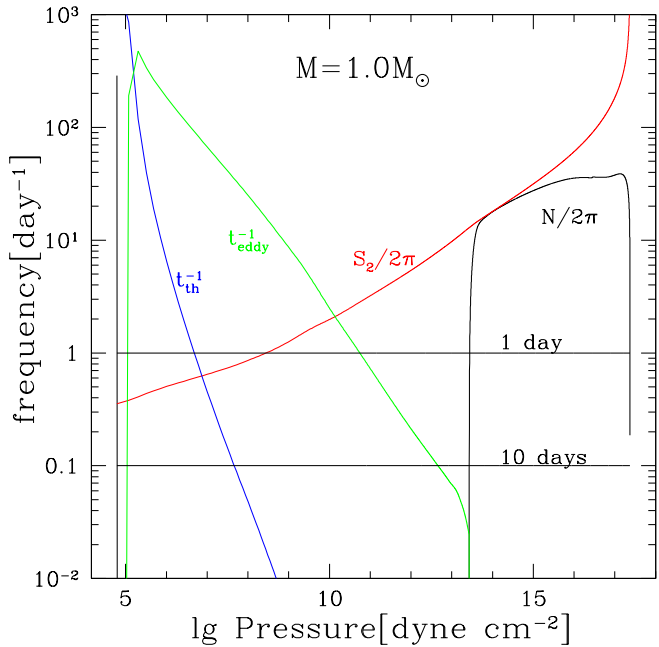,angle = 0,width = \linewidth}}
\caption{Important oscillations frequencies and time scales as a
  function of pressure for a $1\msun$ main-sequence star.  The four
  curves show the Brunt-V\"ais\"all\"a frequency $N$ (black), Lamb
  frequency $S_l$ (red; $\ell = 2$ is shown), inverse thermal time
  $t_{\rm th}^{-1}$ (blue), and inverse eddy turnover time
  $t^{-1}_{\rm ed}$ (green).  Large, real values of $N$ occur in the
  radiative core and very near the photosphere, while $N^2<0$ in the
  convective envelope.  Gravity waves propagate only where the angular
  frequency is below both $N$ and $L_\ell$.  The two horizontal lines
  delimit the range of tidal forcing frequencies of interest here.}
  \label{fig:prop_1.0msun}
\end{figure}

The equilibrium analysis ignores fluid inertia and the excitation of
the star's natural oscillation modes.  While this assumption may be
valid near the surface of the star, it does not hold deeper in the
interior.  Gravity waves ($g$-modes; restored by buoyancy) can
propagate in the radiative interiors of Sun-like stars with a range of
oscillation periods that includes the tidal forcing periods of
interest ($\la$3 days).  Tidal forcing of radiative regions may
produce substantial deviations from hydrostatic balance, as well as
large surface amplitudes of $\Delta F/F$, in particular if resonant
oscillations are excited.  This is especially relevant for
main-sequence stars of mass $M \ga 1.4$--$1.5\msun$ with mainly
radiative envelopes.  Less massive stars ($M \la 1.3$--$1.4\msun$)
have rather deep convective envelopes that can block information about
the dynamic interior from being conveyed to the surface.  Here we
investigate each of these regimes with both analytic estimates and
numerical models of oscillating stars.

Our calculations employ realistic models of 0.9--$1.6\msun$
main-sequence stars, constructed with the EZ stellar evolution code
\citep{Paxton2004}, a distilled and rewritten version of the program
originally created by Peter Eggleton.  We adopt Solar metallicity and
a convective mixing length of 1.6 times the pressure scale height.
All stars are evolved to an age when the core hydrogen abundance has
the Solar value of $X_H=0.35$.  Models with 199 radial grid points are
interpolated to yield $\ga$$10^4$ points in which the $g$-mode radial
wavelength is well resolved in the core.

\begin{figure}
\centerline{\epsfig{file = 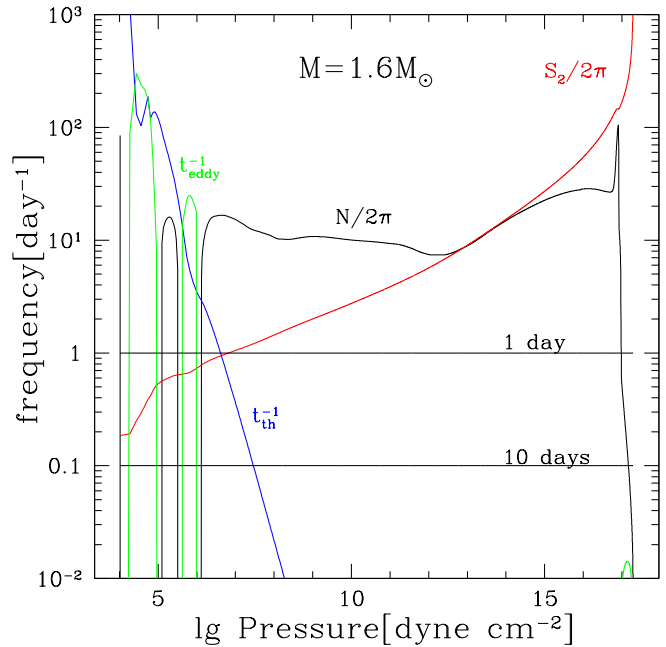,angle = 0,width = \linewidth}}
\caption{ Same as Fig.~\ref{fig:prop_1.0msun}, but for a $1.6M_\odot$
  main-sequence star. Note two geometrically thin, relatively
  inefficient ($t_{\rm th} \sim t_{\rm ed}$) convection zones near the
  surface. The spike in $N$ near the center is at the edge of the
  convective core, and signals a steep gradient in the mean molecular
  weight.}
  \label{fig:prop_1.6msun}
\end{figure}

Figures~\ref{fig:prop_1.0msun} and \ref{fig:prop_1.6msun} illustrate
some of the differences between $1\msun$ and $1.6\msun$ stars, and
serve to introduce several important physical quantities used in the
remainder of this section.  The Lamb frequency,
\be\label{eq:Lamb}
S_\ell = [\ell(\ell + 1)]^{1/2}\frac{c_s}{r}~,
\ee
is the inverse of the horizontal sound-crossing time scale, where $c_s$ 
is the sound speed, and $[\ell(\ell + 1)]^{1/2}/r \equiv k_h$ is the 
horizontal wavenumber of the oscillation.  For fixed chemical 
composition, the squared Brunt-V\"ais\"all\"a frequency is
\be\label{eq:Brunt}
N^2 \simeq \frac{g}{H_p}(\nabla _{\rm ad} - \nabla)~,
\ee
where $H_p = -(d\ln p/dr)^{-1} = p/(\rho g)$ is the pressure scale height,
and $\nabla = d\ln T/d \ln p$ is the temperature gradient\footnote{Do
  not confuse the temperature gradient $\nabla$ with the spatial
  gradient $\grad$ used in \S~\ref{sec:eqtide}.}  ($\nabla_{\rm ad}$
is the adiabatic value).  Radiative regions have $\nabla_{\rm ad} - \nabla 
> 0$ ($N^2 > 0$), and $N$ represents the frequency of buoyancy
oscillations.  In convection zones, $\nabla_{\rm ad} - \nabla < 0$ and
$N^2 < 0$, indicating that $g$-modes are evanescent.  When $N^2 < 0$,
the time scale
\be\label{eq:teddy} 
t_{\rm ed} \sim  |N|^{-1}~, 
\ee
approximates the turnover time of convective motions (for details and
modifications for radiative losses, see, e.g., Kippenhahn \& Weigert
1990).  A shell of radius $r$, thickness $H_p$ (size of the largest
convective eddies), and radiative luminosity $L$ cools on the thermal
time scale
\be\label{eq:tth}
t_{\rm th} = \frac{4\pi r^2 H_p \rho C_p T}{L}~,
\ee
where $C_p$ is the specific heat at constant pressure.     

The $1\msun$ model (Fig.~\ref{fig:prop_1.0msun}) has one deep
convection zone with $t_{\rm ed} \ll t_{\rm th}$ over most of the
region, indicating that convection very efficiently transports energy
and causes the zone to be essentially isentropic.  By consrast, the
$1.6\msun$ star (Fig.~\ref{fig:prop_1.6msun}) has two thin surface
convection zones with $t_{\rm ed} \sim t_{\rm th}$, and thus the
radiative and convective fluxes are comparable.  Gravity waves with
frequency $\omega$ propagate only in radiative regions where $\omega <
N$ and $\omega < S_\ell$.  For the $1\msun$ star, heat and entropy
generated by $g$-modes in the radiative interior may be strongly
mitigated owing to the long thermal time at the base of the deep
convection zone.  On the other hand, $g$-modes in a $1.6\msun$ star
can propagate very near the surface, producing qualitatively different
results.

We now go on to elucidate the physics of the flux perturbations.  All
the analytic and numerical work that follows assumes that the tidal
potential has the generic form $U \propto r^\ell Y_{\ell m}(\theta,
\phi) \exp (-i\omega t)$ with forcing frequency $\omega$.

\begin{figure}
\centerline{\epsfig{file = 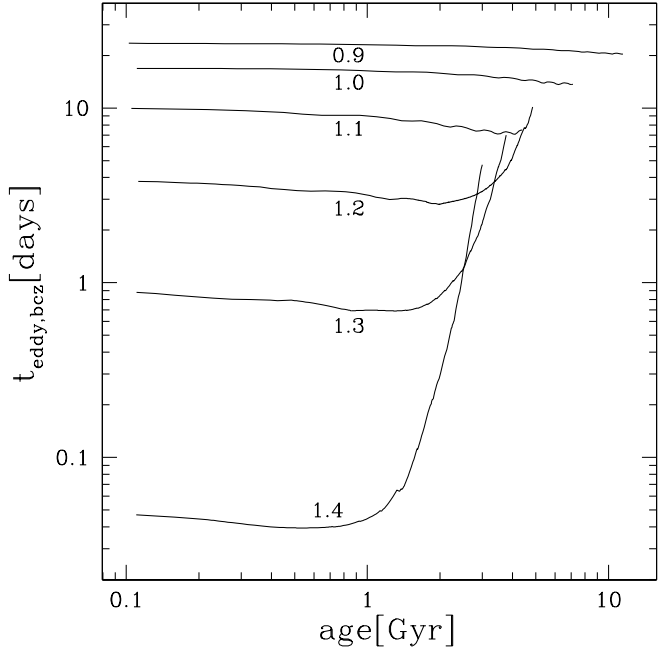,angle = 0,width = \linewidth}}
\caption{Eddy turnover time at base of the convective envelope versus
  stellar age for a range of stellar masses.}
\label{fig:ted_vs_age}
\end{figure}
%

% -----------------------------------------------------------

\subsection{ Heat Transfer in a Convective Envelope }
\label{sec:conv}

Calculation of the perturbed convective flux in oscillating stars is a
thorny issue.  For the purposes of our study, we argue for an
especially simple treatment that draws from previous work on this
subject.  Specifically, we modify the prescription of Brickhill (1983,
1990; see also Goldreich \& Wu 1999a,b), which was originally applied
to white-dwarf pulsations, into a form appropriate for the tidal flow
problem.

In the mixing-length theory of convection, heat is transported by
eddies with a spectrum of sizes $l\la H_p$, speeds $v_l$, and turnover
times $t_{\rm ed}(l) = l/v_l$.  The Kolmogorov scalings for turbulent
motions give $v_l \propto l^{1/3}$, $t_{\rm ed} \propto l^{2/3}$, and
an energy density per unit mixing length interval $\propto$$l^{-1/3}$.
We see that in the unperturbed star most of the convective energy flux
($\propto$$v_l^3$ at scale $l$) is carried by the largest eddies ($l
\sim H_p$).  Convection efficiently transports energy when the
radiative thermal time scale associated with the dominant eddies is
much longer than $t_{\rm ed}$.  Alternatively, efficient convection
implies that the gradient of the specific entropy $s$ is small; i.e.,
$d\ln s/d \ln p \ll 1$.  If all the convective energy flux $F$ is
carried by eddies with mixing length $l$, the flux and entropy
gradient are related by \citep[e.g.,][]{Kippenhahn1990}
\be\label{eq:dsdlnp}
\frac{1}{C_p}\frac{ds}{d\ln p}(l) & = & \left( \nabla -
  \nabla_{\rm ad} \right) \sim \left[ \frac{F}{p c_s} \left(
    \frac{H_p}{l} \right)^2 \right]^{2/3}~.
\ee
Efficient convection enforces $\nabla - \nabla_{\rm ad}\ll 1$, which
implies $d\ln s/d \ln p \ll 1$, since $s \ga C_p$ in the convective
regions of our background models.

\begin{figure}
 \centerline{\epsfig{file = 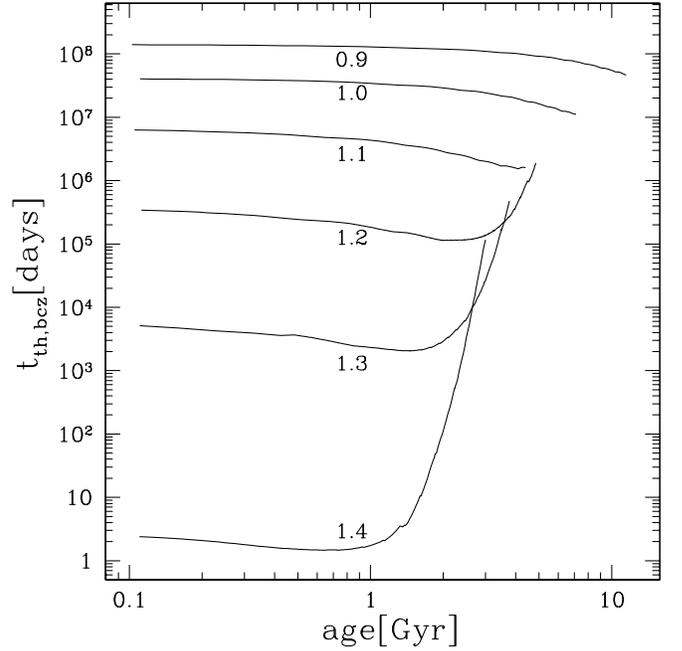,angle = 0,width = \linewidth}}
 \caption{Thermal time at the base of the convection envelope versus
   stellar age for a range of stellar masses.}
 \label{fig:tth_vs_age}
\end{figure}

Gravity waves with the tidal forcing frequency $\omega$ are excited in
the radiative region below the convection zone.  Convective eddies can
transport heat during a forcing period only if $t_{\rm ed}< 2\pi/\omega$
\citep[e.g.,][]{Brickhill1990,Goldreich1999b}.  Inspection of Fig.~
\ref{fig:prop_1.0msun} shows that in the $1\msun$ model, the largest
eddies have $t_{\rm ed} \simeq 20 (p/p_{\rm bcz})^{0.5}$\,days, where
$p_{\rm bcz} \simeq 10^{13.5}\ {\rm dyne\ cm^{-2}}$ is the pressure at
the base of convection zone.  Using the Kolmogorov scaling, the
``resonant'' length $l_{\rm res}$ for which $\omega t_{\rm ed}/2\pi =
1$ is
\be\label{eq:lres}
\frac{l_{\rm res}}{H_p} &
\sim & 10^{-2}\ \left( \frac{2\pi/\omega}{1\ {\rm day}} \right)^{3/2}
\left( \frac{p_{\rm bcz}}{p} \right)^{3/4}~,
\ee 
which is $>$1 for all periods $2\pi/\omega > 1$\,day when $p \la
10^{12} \ {\rm dyne\ cm^{-2}}$, which still encompasses much of the
convection zone.  Now imagine the situation where all the
convective flux is carried by eddies of size $\la$$l_{\rm res}$.  The
entropy gradient for this range of mixing lengths is
\be\label{eq:entropyres}
 \frac{1}{C_p} \frac{ds}{d\ln p}(l_{\rm res})
& \sim & 10^{-3} 
\left(\frac{2\pi/\omega}{1\ {\rm day}}\right)^{-1}
 \left( \frac{p}{p_{\rm bcz}} \right)~,
\ee 
where we have adopted $F/pc_s \sim 10^{-8}$ at the base of the
convection zone, as indicated by our $1\msun$ stellar model.

These arguments suggest that convection is efficient in a $1\msun$
star at the forcing periods of interest even if small ``resonant''
eddies carry all the energy flux near the base of the convection zone.
At larger radii, but not too near the photosphere, convection is both
efficient and rapid ($\omega t_{\rm ed}/2\pi < 1$) over the full
spectrum of eddies.  Rapid convection on all scales $l \la H_p$
enforces isentropy in the convection zone, such that $s$ and its
Lagrangian perturbation $\Delta s$ are nearly constant, as in the
\citet{Brickhill1983,Brickhill1990} picture.  While convection at the
base is rapid only on small scales, it is still highly efficient,
which yields $s \simeq \textrm{constant}$ and further indicates that
$\Delta s/C_p$ is small in magnitude, as we demonstrate in
\S~\ref{sec:thickconv}.

As the stellar mass increases, the convection zone thins and
$t_{\rm ed}$ at the base decreases (see Figs.~\ref{fig:prop_1.6msun}
and \ref{fig:ted_vs_age}).  Rapid convection holds over the
bulk of the convection zone for masses
$\ga$$1\msun$.  However, the assumption that the convection is
efficient starts to break down at 1.4--$1.5\msun$, since $t_{\rm th}
\sim t_{\rm ed}$ at the base (see
Figs.~\ref{fig:prop_1.6msun} and \ref{fig:tth_vs_age}).  For the full
range of stellar masses considered here, we assume that $s$ and
$\Delta s$ are constant in convection zones.  

% -----------------------------------------------------------

\subsection{Analytic Result for Thick Convection Zones}
\label{sec:thickconv}

In a fully convective star, the emergent luminosity is determined
entirely by the surface boundary conditions.  Under our assumption
that $\Delta s$ is constant in the convection zone, the perturbed
luminosity is likewise a function only of the boundary conditions.
Stars of mass $\la$1.3--$1.4\msun$ have long thermal times ($\omega
t_{\rm th} \gg 1$) at the top of the interior radiative region (see
Fig. \ref{fig:tth_vs_age}), so that the flux perturbation $\Delta F$
is approximately the ``quasi-adiabatic'' value, derived by ignoring
$\Delta s \propto (\omega t_{\rm th})^{-1}$ in eq.~(\ref{eq:y5}).  We
assume efficient convection continues to just below the photosphere.

At the photosphere, we adopt the usual Stefan-Boltzmann relation, $F =
\sigma T^4$, and the hydrostatic condition, $p\kappa/A=2/3$, where $A$
is the total acceleration defined in \S~\ref{sec:vonzeipel}, and 2/3
is the photospheric optical depth.  Taking the photosphere to define
the stellar surface, we compute the Lagrangian perturbations,
\be\label{eq:dphotf}
\frac{\Delta F}{F} = 4\frac{\Delta T}{T}~,
\ee
and 
\be\label{eq:dphotp} 
\frac{\Delta p}{p} - \frac{\Delta A}{g} + \frac{\Delta \kappa}{\kappa} = 0~.
\ee
Using $s$ and $p$ as our independent thermodynamic variables, we write
$\Delta \kappa/\kappa=\kappa_{\rm ad}\Delta p/p + \kappa_s \Delta
s/C_p$ and $\Delta T/T=\Delta s/C_p + \nabla_{\rm ad}\Delta p/p$.
In our numerical work (see \S~\ref{sec:numerical}), we self-consistently compute the perturbation $\Delta A$ to the effective surface gravity, in order to follow resonant oscillations, where the equilibrium-tide result fails.  However, we are now addressing non-resonant forcing, for which 
we use the equilibrium-tide approximation at the surface, giving
$\Delta A/g=-\partial \xi_r/\partial r$ (see \S~\ref{sec:vonzeipel}).
We now have
\be
\frac{\Delta p}{p} & = & - \left(\frac{\kappa_s \Delta s/C_p + 
\partial\xi_r/\partial r}{1+\kappa_{\rm ad}} \right)~,
\ee
and upon substitution, 
\be\label{eq:surfacedf}
\frac{\Delta F}{F} = 
 4 \left( \frac{1 + \kappa_{\rm ad}- \nabla_{\rm ad} \kappa_s}{1 + \kappa_{\rm ad}} \right)
  \frac{\Delta s}{C_p}
- \frac{4\nabla_{\rm ad}}{1 + \kappa_{\rm ad}} \frac{\partial\xi_r}{\partial r}~.
\ee
Equation (\ref{eq:surfacedf}) differs from \citet{Goldreich1999a} in
that we retain the gravity perturbation in eq.~(\ref{eq:dphotp}),
whereas they consider a constant-gravity atmosphere (and no tidal
perturbation).  For $g$-modes in white dwarfs, the interesting region
is near the surface and the motion is mainly horizontal, so that
$\Delta A = \Delta g = 0$ is a good approximation. Since the
equilibrium tide has large vertical motions, the $\Delta A$ term must
be retained.

The luminosity change across the convection zone is derived from the
entropy equation (eq.~[\ref{eq:y6}]).  If we ignore horizontal flux
perturbations (set $\ell = 0$ in eq.~[\ref{eq:y6}]) and energy
generation, the equation for the luminosity perturbation $\Delta
L/L=2\xi_r/r + \Delta F/F$ is
\be\label{eq:entropy}
\frac{d(\Delta L/L)}{dM_r} & = & i\omega T \Delta s/L~.
\ee
Integrating over the convection zone with constant $\Delta s$, we
obtain
\be\label{eq:ldiff}
\frac{\Delta L_{\rm ph}}{L} - \frac{\Delta L_{\rm bcz}}{L} = 
i\omega \Delta s \int_{\rm cz} dM_r T/L~, 
\ee
where the subscript ``ph'' refers to the photosphere.  We define
$t_{\rm cz} = C_{p,{\rm ph}} \int_{\rm cz} dM_r T/L$ to be the mean
thermal time of the convection zone, so that the right-hand side of
eq.~(\ref{eq:ldiff}) is $i\omega t_{\rm cz} \Delta s/C_{p,{\rm ph}}$.

Figure~\ref{fig:tth_vs_age} shows that the thermal time at the base of
the convection zone (of order $t_{\rm cz}$) for $M \la 1.3\msun$ is
orders of magnitude longer than the forcing periods of 1--10\,days.
Insofar as $|\Delta L|/L \sim |\xi_r|/r$ at any location in the star
(i.e., if resonances are neglected), we see that
$(|\xi_r|/r)^{-1}|\Delta s|/C_p \sim (\omega t_{\rm cz})^{-1} \ll 1$
in stars with deep convective envelopes.  In this limit,
eq.~(\ref{eq:surfacedf}) becomes
\be\label{eq:simplesurfacedf}
\frac{\Delta F}{F} \simeq
- \frac{4\nabla_{\rm ad}}{1 + \kappa_{\rm ad}} 
\frac{\partial\xi_r}{\partial r}~.
\ee
If we had set $\Delta A = 0$, the amplitude of the photospheric flux
perturbation would have been $\sim$$|\Delta s|/C_p$ rather than the
much larger value $\sim$$|\xi_r|/R$.

Photospheric flux perturbations in tidally forced solar-type stars
with thick convective envelopes arise mainly from changes in the local
effective gravity.  This statement is reminiscent of, but physically
distinct from, von Zeipel's theorem (eqs.~[\ref{eq:fluxvar}] and
[\ref{eq:fluxvar2}]).  We have recovered our equilibrium-tide scaling,
$\Delta F_\ell/F = -\lambda_\ell \xi_{r,\ell}/R$, where
eq.~(\ref{eq:simplesurfacedf}) gives
\be\label{eq:lambda}
\lambda_\ell = 4(\ell + 2) 
\frac{\nabla_{\rm ad}}{1 + \kappa_{\rm ad}}~.
\ee
For $M = 1.0$--$1.4\msun$, we find $\lambda_2 \simeq 1.9$--1.1.  These
estimates neglect resonant excitation of $g$-modes, a point
addressed in \S~\ref{sec:numerical}.
 
% -----------------------------------------------------------

\subsection{ Analytic Result for Radiative Envelopes }
\label{sec:radiative}

As the stellar mass increases beyond $1.4\msun$, the outer convective
region thins and sits close to the surface, where $t_{\rm ed} \sim
t_{\rm th}$.  Figure~\ref{fig:prop_1.6msun} shows that the $1.6\msun$
model has two thin, inefficient surface convection zones, as well as a
convective core.  Radiative energy transport is important throughout
the envelopes of these more massive stars.  We now consider the
idealized case of a completely radiative envelope, and obtain an
analytic approximation for $\Delta L/L$ at the surface.

Near the surface of a radiative star, we have $H_p/r \ll 1$, $4\pi r^3
\rho /M_r \ll 1$, and $\omega^2 r/g \ll 1$ for $2\pi/\omega =
1$--10\,days.  Under these conditions, the quasi-adiabatic luminosity
perturbation becomes \citep[e.g.,][]{Unno1989}
\be
\frac{\Delta L_{\rm qad}}{L} \simeq 
- \zeta \frac{\Delta p}{p} 
 + \frac{gk_h^2H_p}{\omega^2} 
 \left( \frac{\nabla_{\rm ad}}{\nabla} - 1 \right)
 \left( \frac{\Delta p}{p}+ 
\frac{\xi_r-\xi_{r,\rm eq}}{H_p} \right) \nonumber \\
\label{eq:qaddl}
\ee
where 
\be
\zeta=\kappa_{\rm ad} - 4\nabla_{\rm ad} + \frac{\nabla_{\rm ad}}{\nabla}
- \frac{d\nabla_{\rm ad}}{d\ln T}~,
\ee
and $\xi_{r,\rm eq}$ is the equilibrium-tide radial displacement
(eq.~[\ref{eq:eqtide}]).  Nonzero values of $\Delta p/p$ and $(\xi_r -
\xi_{r,{\rm eq}})/H_p$ indicate deviations from hydrostatic
equilibrium.  Care must be taken with these terms, because the
denominators $p$ and $H_p$ become very small close to the surface.

With the help of the Appendix, we define the variables
\be
\alpha & = & y_1-y_2+y_3=
-\frac{H_p}{r}\frac{\Delta p}{p}~, \\
\beta & = & y_2+\frac{U}{gr}=
\frac{H_p}{r}\frac{\Delta p}{p} + \frac{\xi_r-\xi_{r,\rm eq}}{r}~,
\ee
which satisfy the differential equations
\be\label{eq:dalphadr}
\frac{d\alpha}{d r} & \simeq & - \frac{d\ln \rho}{d r} \alpha
+ \frac{gk_h^2}{\omega^2} \beta + (\ell+4)\frac{U}{gr^2}~, \\
\frac{d\beta}{d r} & = & - \frac{N^2}{g} \alpha + \frac{\beta}{r}~.
\ee
When $\omega^2 \ll gk_h^2 H_p$, these equations produce the $g$-mode
dispersion relation $k_r^2=k_h^2N^2/\omega^2$ (in the limit
$k_r^2/k_h^2 \ll 1$) for radial wavenumber $k_r$. For these
propagating waves, the surface amplitudes of $\alpha$ and $\beta$ are
determined at the core radiative-convective boundary, where $g$-modes
are driven \citep[e.g.,][]{Goldreich1989}.  On the other hand, when
$\omega^2 \gg g k_h^2 H_p$, the $g$-modes are evanescent
\citep[see][]{Unno1989} and we neglect the term $gk_h^2\beta/\omega^2$
in eq.~(\ref{eq:dalphadr}).  This limit yields the approximate
solution $\alpha \simeq - (\ell+4)(H_p U/gR^2)$, or $\Delta p/p \simeq
(4+\ell)U/gR$.  In this case, $\Delta p/p$ is not small compared to
the fractional fluid displacement, and thus the equilibrium-tide
approximation loses validity.

\begin{figure}
\centerline{\epsfig{file = 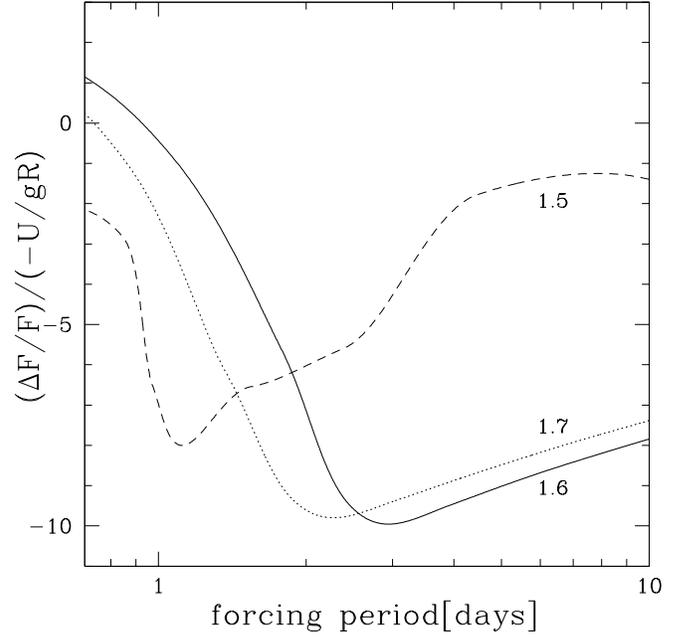,angle = 0,width = \linewidth}}
\caption{ Ratio of surface Lagrangian flux perturbation $\Delta F/F$
  to equilibrium-tide displacement $-U/gR$ for a range of forcing
  periods in the limit where surface $g$-modes are evanescent.  The
  flux is evaluated at the location where $\omega t_{\rm th} = 1$.
  Dashed, solid, and dotted curves correspond to $M=1.5$, 1.6, and
  $1.7\msun$, respectively.} \label{fig:radiativeflux}
\end{figure}

From our stellar models, we find that the evanescent regime
corresponds to forcing periods of $\la$4--8\,days for $M =
1$--$1.6\msun$, most of the range of interest.  The high-frequency
limit of eq.~(\ref{eq:qaddl}) is
\be\label{eq:dlqadapprox}
\frac{\Delta L_{\rm qad}}{L} \simeq -\zeta (\ell+ 4)\ \frac{U}{gR}~.
\ee
This relation should be evaluated at the layer where $\omega t_{\rm
  th} \simeq 1$, above which the luminosity effectively ``freezes
out.''  Figure \ref{fig:radiativeflux} shows the quasi-adiabic flux
perturbation $\Delta F_{\rm qad}/F=\Delta L_{\rm qad}/L-2\xi_r/R$,
evaluated where $\omega t_{\rm th}=1$, for a range of forcing periods
and $M = 1.5$--$1.7\msun$. Note that $|\Delta F/F|$ can be an order of
magnitude larger than $|U|/gR$, because of the rather large values of
$|\zeta|(\ell + 4)$ for $\ell \geq 2$.  Much larger perturbations are
possible when $g$-modes are resonantly excited in a radiative star, as
we discuss in the next section.

We must point out that the quasi-adiabatic approximation is
technically inappropriate when $\omega t_{\rm th} \sim 1$.
Equation~(\ref{eq:dlqadapprox}) should be viewed as an estimate of the
modulus of the luminosity perturbation at the surface.  If, for
instance, $|\Delta s|/C_p \ga |U|/gR$ where $\omega t_{\rm th} \sim
1$, then $\Delta L/L$ at the surface will have a substantial imaginary
part (see eq.~[\ref{eq:entropy}]).  This is what we find in the
numerical calculations summarized in the next section.

%

% -----------------------------------------------------------

\subsection{ Numerical Examples }
\label{sec:numerical}

\begin{figure*}
\centerline{\epsfig{file = 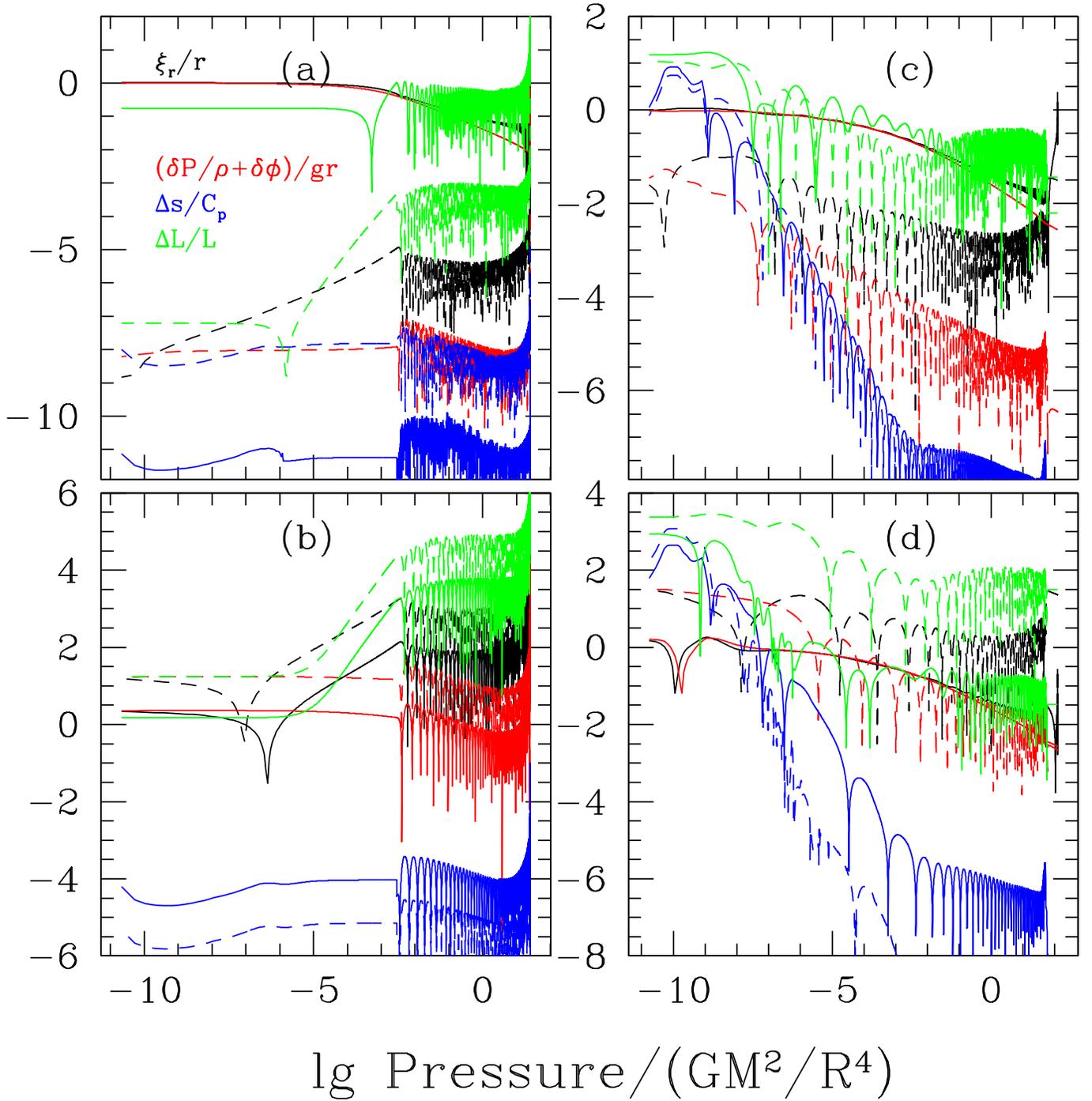,angle = 0,width = 1.\linewidth}}
\caption{Responses of tidally forced $1\msun$ and $1.6\msun$
  main-sequence stars.  Black, red, blue, and green curves denote,
  respectively, the logarithms of $\xi_r/r$, $(\delta p+\rho \delta
  \varphi)/\rho g r$, $\Delta s/C_p$, and $\Delta L/L$.  Solid
  (dashed) curves show the real (imaginary) parts.  The tidal
  potential has been scaled so that $\xi_r/R = 1$ corresponds to the
  equilibrium-tide value.  The four panels show the following: (a)
  non-resonant response of a $1\msun$ star tidally forced at a period
  of $2\pi/\omega \simeq 2.91\,{\rm day}$; (b) resonant response of a
  $1\msun$ star with $2\pi/\omega \simeq 1.00\,{\rm day}$; (c)
  non-resonant response of a $1.6\msun$ star with $2\pi/\omega \simeq
  3.00\,{\rm day}$; (d) resonant response of a $1.6\msun$ star with
  $2\pi/\omega \simeq 1.02\,{\rm day}$}
  \label{fig:allvsdepth}
\end{figure*}

Here we show solutions of the perturbed mass, momentum, and energy
equations that describe linear, nonadiabatic oscillations of a star
subject to a varying tidal force.  The equations listed in the
Appendix are the same as in \citet{Unno1989} for radiative
regions, but augmented to include the tidal acceleration.  
In convection zones, we apply the prescription $\Delta s =
\textrm{constant}$ based on our conclusions in \S~\ref{sec:conv}.
Figure~\ref{fig:allvsdepth} summarizes how the interiors of $1\msun$
and $1.6\msun$ stars respond to resonant and non-resonant tidal
forcing.  The tidal potential has been scaled so that $\xi_r/R = 1$
corresponds to the equilibrium-tide surface displacement.

For our $1\msun$ model, the non-resonant response to a forcing period
of $\simeq$3\,days is shown in Fig.~\ref{fig:allvsdepth}a.  We see
that $\xi_r/R$ matches the equilibrium-tide result at the surface; the
imaginary piece is completely negligible.  We also find that our
approximation for $\Delta F/F$ at the surface
(eq.~[\ref{eq:simplesurfacedf}]) works very well.  A factor of
$\sim$10 decay in $|\Delta L|/L$ occurred in order for $|\Delta F|/F
\sim \xi_r/R$ at the surface.  Variation of $\Delta s/C_p$ in the
convection zone ($\log [p/(GM^2/R^4)] > -2.5$) is due to changes in
$C_p$.  In the radiative interior, the oscillations are caused by most
nearly resonant $g$-modes, whose amplitudes rise rapidly as the core
is approached, due to conservation of wave luminosity.  We have
checked that the quasi-adiabatic approximation of $\Delta L/L$ is
valid in the radiative region; the ratio of the real and imaginary
parts is found to be roughly constant for the ingoing gravity-wave
\citep[see also][]{Zahn1977}.

\begin{figure}
\centerline{\epsfig{file = 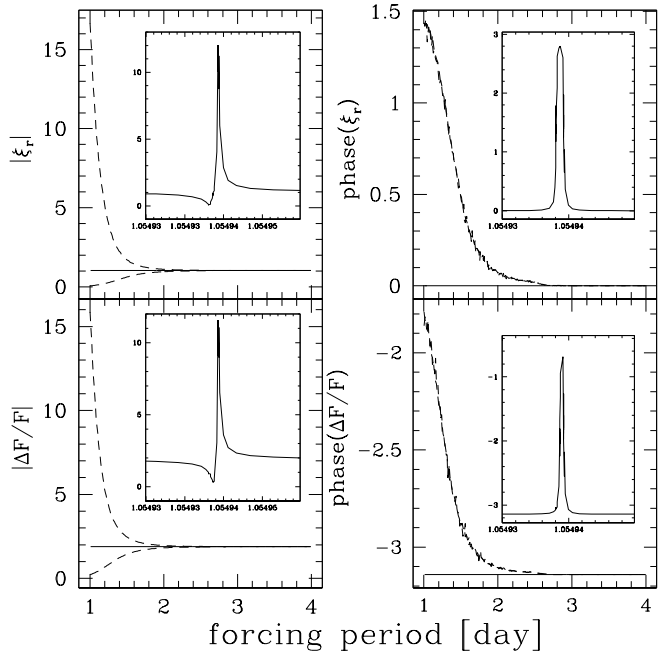,angle = 0,width = \linewidth}}
\caption{ Surface radial displacement and Lagrangian flux perturbation
  versus forcing period for a $1.0M_\odot$ star. Solid lines connect
  points halfway between resonant $g$-mode periods, while dashed
  curves give the maximum and minimum values found on resonance.  The
  equilibrium-tide approximation is extremely good, except when the
  forcing period is $<$2\,days and resonant.}
\label{fig:xirdf_magphase_env_1.0msun}
\end{figure}

In order to model the resonant response of a $1\msun$ star, we tuned
the forcing period to $\simeq$1\,day (see Figs.~\ref{fig:allvsdepth}b
and \ref{fig:xirdf_magphase_env_1.0msun}).  At the surface, both
$\xi_r$ and $\Delta L$ have dominant imaginary parts, due to the short
radial wavelength of the $g$-mode compared to the equilibrium-tide
fluid displacement.  The entropy at the base of the convection zone is
very strongly perturbed in comparison to the non-resonant case, but
$\Delta L$ is still damped by orders of magnitude as the surface is
approached.

Figure \ref{fig:xirdf_magphase_env_1.0msun} shows the surface values
of the complex modulus and phase of $\xi_r/R$ and $\Delta F/F$ versus
forcing period.  The phase is $\tan^{-1}(\textrm{Imaginary/Real}) \in
(-\pi, \pi)$.  Solid lines connect points halfway between $g$-mode
resonant periods.  We find that the equilibrium-tide approximation
given by eqs.~(\ref{eq:eqtide}) and (\ref{eq:simplesurfacedf}) is
excellent for non-resonant forcing.  Dashed curves give the maximum
and minimum values that occur on resonance.  One example of a
resonance is shown in the insets.  Resonant forcing at periods of
$<$2\,days yields surface values of $\xi_r/R$ and $\Delta F/F$ that
differ substantially from the equilibrium-tide results.  However, the
ratio of resonance width to the spacing between adjacent resonances is
$\sim$$10^{-4}$, making resonant forcing very unlikely.  It is
noteworthy that at forcing periods of $>$2\,days, the equilibrium-tide
result holds extremely well even when precisely on a resonance.  As
explained by Zahn (1975), the resonant response can be considered as
the sum of the equilibrium tide and the most nearly resonant wave.  As
the period increases, the $g$-mode radial wavelength decreases,
resulting in a reduction of the overlap integral for the mode and the
tidal force, which in turn gives a decreased amplitude of the wave
component relative to the equilibrium tide.

The non-resonant response of the $1.6\msun$ star is shown in
Fig.~\ref{fig:allvsdepth}c.  We see that the equilibrium-tide result
provides a good match to $\xi_r/R$.  Our estimate for the modulus of
the radiative luminosity perturbation in the evanescent limit
(eq.~[\ref{eq:dlqadapprox}]) agrees reasonably well with what is in
Fig.~\ref{fig:allvsdepth}c.  We also see that $\Delta L/L$ does
roughly ``freeze-out'' when $\omega t_{\rm th} \simeq 1$, just below
the base of the convection zone at $\log p/(GM^2/R^4) \simeq -9$ (see
Fig.~\ref{fig:prop_1.6msun}).  Our expectations in
\S~\ref{sec:radiative} regarding the imaginary part of $\Delta L/L$
are borne out in Fig.~\ref{fig:allvsdepth}c

A resonantly excited $1.6\msun$ star exhibits huge surface flux
perturbations, radial displacements, and phase lags, as seen in
Fig.~\ref{fig:allvsdepth}d.  In
Fig.~\ref{fig:xirdf_mag_point_1.6msun}, surface values of $|\xi_r|/R$
and $|\Delta F|/F$ are plotted as a function of forcing period, where
we have taken care to resolve resonances.  Resonant amplitudes vary
non-monotonically with period, in contrast to the smooth behavior of
the $1\msun$ star (Fig.~\ref{fig:xirdf_magphase_env_1.0msun}).
Although we do not show the results here, similar plots for masses
between $1\msun$ and $1.6\msun$ show progressively more structure as
the mass increases.  The cause of this irregularity is not clear, but
may have to do with the two thin surface convection zones changing the
overlap of successive $g$-modes with the tidal force.

%

% -----------------------------------------------------------

\section{SUMMARY}
\label{sec:summary}

We have investigated in detail the ellipsoidal oscillations of
0.9--$1.6\msun$ main-sequence stars induced by substellar companions.
Classical models of ellipsoidal variability
\citep[e.g.,][]{Wilson1994} are built on the assumption of hydrostatic
balance in a frame corotating with the binary orbit.  This approach is
justified in the context of short-period ($P_{\rm orb} \la 10$\,days)
binaries containing two stars of comparable mass, where tidal
dissipation circularizes the orbits and synchronizes the stellar spins
with the orbital frequency.  However, when the companion has a very
low mass, we cannot assume that the binary is in complete tidal
equilibrium; in fact, this state may be unattainable (see
\S~\ref{sec:prelim}).  In this case, one must, in general, appeal to a
dynamical description of the tidal interaction.  A substellar
companion with $P_{\rm orb} \ga 1$\,day raises tides on the star that
are a small fraction of the stellar radius (see
eq.~[\ref{eq:tidescale}]), permitting a linear analysis of the stellar
oscillations.

While the root of our study is a dynamical treatment of stellar tidal
perturbations, the equilibrium-tide approximation does have an
important realm of validity (see below).  For this reason, we derived
in \S~\ref{sec:eqtide} a general expression (eq.~[\ref{eq:fdiskeq}])
for the measurable flux variation of a star that remains in
hydrostatic equilibrium under the influence of a small external tidal
force.  This formula (1) assumes that the local perturbation to the
energy flux at the stellar surface is proportional to and in phase
with the equilibrium-tide radial fluid displacement at each angular
order $\ell$ (eq.~[\ref{eq:eqtide}]), (2) neglects stellar rotation,
and (3) applies to inclined and eccentric orbits.  As expected, the
fractional amplitude of the modulation is $\sim$$\pert \equiv
(M_p/M)(R/a)^3$ for small eccentricities and $I = 90^\circ$, or
$\sim$$10^{-5} (M_p/M_J)(P_{\rm orb}/1\,{\rm day})^{-2}$ for a star
like the Sun (see \S~\ref{sec:prelim}).

\begin{figure}
\centerline{\epsfig{file = 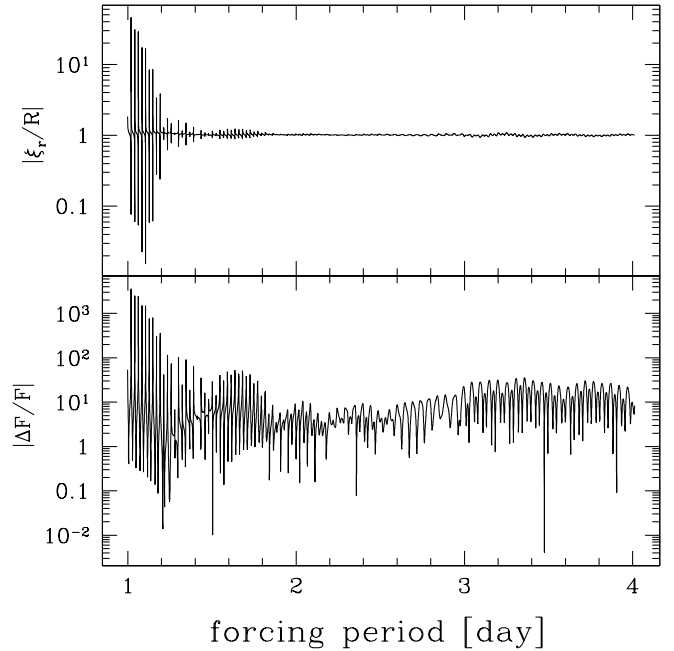,angle = 0,width = \linewidth}}
\caption{ Surface radial displacement and Lagrangian flux perturbation
  versus forcing period for a $1.6M_\odot$ star.  Curves connect
  evenly spaced points away from resonances, with finer spacing near
  resonance periods.}
\label{fig:xirdf_mag_point_1.6msun}
\end{figure}

A common practice is to use von Zeipel's theorem when computing the
surface radiative flux from a tidally distorted star (see
\S~\ref{sec:vonzeipel}).  The theorem assumes that the star is in
hydrostatic equilibrium and that the energy transport in the outer
layers is purely by radiative diffusion.  As already mentioned, the
hydrostatic assumption is technically unjustified for substellar
perturbers.  Moreover, the majority of {\em Kepler} targets will be
main-sequence stars with masses of $<$$1.4\msun$, which have
substantial surface convection zones.  Evidently, von Zeipel's theorem
is an inappropriate starting point for the conditions of interest.

Section~\ref{sec:conv} discusses heat transport in perturbed stars
with convective envelopes.  Heuristic arguments are used to develop a
simple treatment of the perturbed convection zone in main-sequence
stars of mass $<$$1.6\msun$ with forcing periods of 1--10\,days.  We
suggest that both the specific entropy $s$ and its Lagrangian
perturbation $\Delta s$ are spatially constant in convective regions,
a model partly inspired by the ideas of
\citet{Brickhill1983,Brickhill1990}.

\input{tab2.tex}

Using this prescription, we analytically compute in
\S~\ref{sec:thickconv} the perturbed flux at the photosphere of deeply
convective stars ($M\la 1.4\msun$), where the thermal time scale at
the base of the convection zone is much longer than the forcing
period.  We find that $\Delta s/C_p$ is negligible near the top of the
convection zone, and that the photospheric flux perturbation is
proportional to changes in the effective surface gravity.  Thus, we
recover the equilibrium-tide result, $\Delta F/F = -\lambda_\ell
\xi_r/R$, at the surface, where $\lambda_\ell$ depends on the
adiabatic derivatives of opacity and temperature with respect to
pressure (see eq.~[\ref{eq:lambda}]).  Numerical solutions of the
equations of linear, nonadiabatic stellar oscillations (see
\S~\ref{sec:numerical} and Fig.~\ref{fig:allvsdepth}a) corroborate our
analytic estimates in the non-resonant regime.  Resonant excitations
of $g$-modes in the radiative stellar interior cause large departures
from the equilibrium-tide approximation when the forcing period is
$<$2\,days (Figs.~\ref{fig:allvsdepth}b and
\ref{fig:xirdf_magphase_env_1.0msun}).  However, the likelihood of
being on a resonance is small, and at periods of $>$2\,day the
equilibrium-tide result holds for $M \simeq 1\msun$ even with resonant
forcing.

Stars of mass $\ga$$1.4\msun$ have thin, relatively inefficient
surface convection zones.  Thus, $g$-modes can propagate very close to
the surface and produce large flux perturbations and fluid
displacements.  Analytic arguments in \S~\ref{sec:radiative} indicate
that the surface flux perturbations in these stars have non-resonant
amplitudes of $\sim$$10\pert$ (eq.~[\ref{eq:dlqadapprox}] and
Fig.~\ref{fig:radiativeflux}), in rough agreement with our numerical
calculations (Fig.~\ref{fig:allvsdepth}c).  As seen in
Figs.~\ref{fig:allvsdepth}d and \ref{fig:xirdf_mag_point_1.6msun}, a
resonantly forced $1.6\msun$ star can exhibit flux perturbation
amplitudes of $>$$100\pert$ at forcing periods of $\simeq$1\,day.
While the amplitudes are not as extreme at longer periods, their
dependence on period is rather erratic
(Fig.~\ref{fig:xirdf_mag_point_1.6msun}), an issue that deserves
further study.  It will be difficult to derive physical
interpretations from the ellipsoidal variability of these more massive
stars.

% -----------------------------------------------------------

\section{DETECTION PROSPECTS}
\label{sec:detection}

The dominant sources of periodic variability of a star with a
substellar companion are transit occultations (when $|\cos I| <
[R+R_p]/a$), Doppler flux modulations, reflection of starlight from
the companion, and ellipsoidal oscillations.  For each of these
signals, Table ~\ref{tab:fluxmod} lists the characteristic amplitude,
period with the largest power in the Fourier spectrum, and orbital
phase(s) at which the light is a maximum or minimum.  The transit
contribution is included for completeness, but its duration is
sufficiently short---a fraction $\simeq$$(R+R_p)/(\pi a)$ of $P_{\rm
  orb}$---that it should often be possible to excise it from the data
\citep[see][]{Sirko2003}.  Of the remaining signals, the Doppler
variability is the simplest, being purely sinusoidal with period
$P_{\rm orb}$ when the orbit is circular.  The dominant $\ell = 2$
piece of the equilibrium-tide approximation to the ellipsoidal
variability (see eqs.~[\ref{eq:fdiskeq}] and [\ref{eq:P2}]) is also
sinusoidal when $e = 0$, but with period $P_{\rm orb}/2$.  Reflection
is more problematic, as its time dependence is generally not
sinusoidal and not known a priori.

If the companion scatters light as a Lambert sphere
\citep[e.g.,][]{Seager2000}, the Fourier spectrum of the reflection
variability has finite amplitude at all harmonics of the orbital
frequency $\Omega$, but the amplitude at $2\Omega$ is roughly 1/5 of
the amplitude at $\Omega$.  Therefore, the reflection and ellipsoidal
variability amplitudes may be similar at a frequency of $2\Omega$ when
$\alpha = 0.1$, $M_p \sim M_J$, and $P_{\rm orb} \simeq 1$\,day.
Also, the orbital phase at which the reflected light is a maximum is
distinct from both the Doppler and ellipsoidal cases, further
distinguishing the signals.  However, Lambert scattering is probably
never appropriate in real planetary atmospheres.  Infrared reemission
of absorbed optical light, multiple photon scattering, and anisotropic
scattering typically conspire to narrow the peak in the reflection
lightcurve and lower the albedo, decreasing the prominence of the
reflection signal.  These issues are sensitive to the atmospheric
chemistry and the uncertain details in models of irradiated giant
planets.  For reasonable choices regarding the atmospheric
composition, calculated optical albedos of Jovian planets range from
$<$0.01 to $\simeq$0.5 \citep{Seager2000,Sudarsky2000}.  Recent
photometric observations of HD~209458, the star hosting the
first-detected transiting giant planet ($P_{\rm orb} \simeq
3.5$\,days), constrain the planetary albedo to be $<$0.25
\citep{Rowe2006}.

Detailed lightcurve simulations will be required to say how well the
different periodic signals can be extracted from the data.  This is
beyond the scope of the current study.  We now do the simpler exercise
of isolating the ellipsoidal modulations and assessing when this
effect alone should be detectable.  For a star of apparent visual
magnitude $V$ and an integration time of $T = 6\,{\rm hr}$, {\em
  Kepler}'s photon shot noise is\footnote{An integration time of $T =
  6$\,hr is chosen for convenience; {\em Kepler}'s nominal exposure
  time is 30\,min.  Here we use the $V$-band flux as a reference, but,
  in fact, the {\em Kepler} bandpass is 430--890\,nm, which spans $B$,
  $V$, and $R$ colors.}
\be
\left(\frac{\delta \fdisk}{\fdisk}\right)_{\rm shot}
\sim 10^{-5} 10^{0.2(V - 12)}
\left(\frac{T}{6\,{\rm hr}}\right)^{-1/2}~.
\ee
Instrumental noise should contribute at a similar level
\citep[e.g.,][]{Koch2006}.  If the data is folded at the orbital
period and binned in time intervals $T \ll P_{\rm orb}$, the shot
noise is suppressed by a factor of $\sim$$n_{\rm orb}^{-1/2}$, where
$n_{\rm orb}$ is the number of folded cycles.  After folding 1 year of
continuous photometric data using $T = 6$\,hr, a star with $V <12$
orbited by a giant planet with $P_{\rm orb} \la 3$\,days may have a
fractional shot noise per time bin of $\la$$10^{-6}$.  This is less
than the ellipsoidal amplitude, $(\delta\fdisk/\fdisk)_{\rm ell}$,
when $I$ is not too small.

The actual situation is not so simple when the data spans of weeks or
months, because the intrinsic stochastic variability of the star will
not have a white-noise power spectrum.  Over times of $\la$1\,day, the
Sun shows variability of $(\delta \fdisk/\fdisk)_{\rm int} \sim
10^{-5}$, but the amplitude rises steeply between $\sim$1 and 10 days
to $\sim$$10^{-3}$.  Intrinsic variability tends to be large near the
rotation period of the star, due mainly to starspots.  Low-frequency
variability may not too damaging for the study of ellipsoidal
oscillations induced by planets with $P_{\rm orb} \la 3$\,days, but
more study is needed.

{\em Kepler}'s target list will contain $\simeq$$10^5$ main-sequence
FGK stars with $V = 8$--14.  The statistics of known exoplanets
indicate that 1--2\% of all such stars host a giant planet ($M_p \ga
M_J$) with $P_{\rm orb} < 10$\,days \citep[e.g.,][]{Marcy2005}. Of
these ``hot Jupiters,'' $\simeq$30\% have $P_{\rm orb} =1$--3\,days.
It seems that a maximum of $\sim$$10^3$ {\em Kepler} stars will have
detectable ellipsoidal modulations.  If we neglect intrinsic stellar
variability and consider only shot noise, then many systems with
$P_{\rm orb} \la 3$\,days and $V < 14$ will have signal-to-noise $S/N
> 1$ after $\sim$100 cycles are monitored; this may amount to $>$100
stars.  Obviously, the number drops when we place higher demands on
$S/N$ and include the intrinsic variability.  The results depend
critically on the distributions of $M_p$ and $P_{\rm orb}$.

In order to better estimate the number of stars with potentially
detectable ellipsoidal oscillation, we perform a simple population
synthesis calculation.  Denote the set of star-planet system
parameters by $\vec{P} = \{ M, M_p, P_{\rm orb}, I\}$, and let
$f(\vec{P})d\vec{P}$ be the probability of having a system in the
4-dimensional volume $d\vec{P}$.  We assume that the planetary orbits
are circular and obtain $(\delta\fdisk/\fdisk)_{\rm ell}$ from the
equilibrium-tide estimate in Table~\ref{tab:fluxmod}.  Given the mass
of the star, we compute its absolute $V$ magnitude on the
main-sequence using the approximation \citep[see also][]{Henry1993}
\be
\mathscr{M}_V = 4.8 - 10.3 \log (M/M_\odot) ~,
\ee
which is in accord with the usual mass-luminosity relation
$\log(L/L_\odot) \simeq 4 \log(M/M_\odot)$ for $M \simeq 1\msun$.
With a maximum apparent magnitude of $V_{\rm max} = 14$ for the {\em
  Kepler} targets, the maximum distance of the star is
\be
D_{\rm max} = 10^{1 + 0.2(14 - \mathscr{M}_V)}\pc ~.
\ee
With a certain signal-to-noise threshold $(S/N)_{\rm min}$, there is a
maximum distance $D_d < D_{\rm max}$ to which the ellipsoidal
variability is detectable.  For a spatially uniform population, the
detectable fraction of systems is $(D_d/D_{\rm max})^3$.  Thus, the
net detectable fraction among all systems is
\be
\mathscr{E} = \int d\vec{P} f(\vec{P}) 
\left(\frac{D_d}{D_{\rm max}}\right)^3~,
\ee
an integral over all relevant $\vec{P}$ space.

When the only noise is intrinsic to the star, $N =
(\delta\fdisk/\fdisk)_{\rm int}$ and $S/N$ is independent of distance,
so that $D_{d,{\rm int}} = D_{\rm max}$ when $S/N > (S/N)_{\rm min}$,
and $D_{d,{\rm int}} = 0$ otherwise.  In the case of pure shot noise,
there is a maximum magnitude $V_d$ for which the ellipsoidal
oscillations are detectable:
\be\label{eq:detfrac}
V_d = 5\log \left[\frac{(\delta\fdisk/\fdisk)_{\rm ell}}{\chi\ (S/N)_{\rm min} }\right]~,
\ee
where $\chi \sim 10^{-8.4}(T_6 n_{100})^{-1/2}$ is the value of
$(\delta\fdisk/\fdisk)_{\rm shot}$ for $V = 0$, $T = 6 T_6\,{\rm hr}$,
and $n_{\rm orb} = 100n_{100}$.  The corresponding distance is given
by $\log [D_{d,{\rm shot}}/10\pc] = 0.2(V_d - \mathscr{M}_V)$ if $V_d
< V_{\rm max}$, and is $D_{d,{\rm shot}} = D_{\rm max}$ when $V_d >
V_{\rm max}$.  We take the maximum detectable distance to be $D_d =
\min\{D_{d,{\rm int}}, D_{d,{\rm shot}}\}$.

\input{tab3.tex}

At this point the simplest approach is to assume that the parameters
$\{ M, M_p, P_{\rm orb}, I\}$ are statistically independent and carry
out a Monte Carlo integration to obtain $\mathscr{E}$.  To this end,
we draw $M$ from the \citet{Kroupa1993} initial mass function in the
range of 0.5--$1.5\msun$.  The planetary mass is chosen from the
distribution $f(M_p)\propto M_p^{-x}$ for $M_p = 1$--$10M_J$.
\citet{Marcy2005} find that $x \simeq 1$ when considering all detected
planets; the shape of $f(M_p)$ is not well constrained at $P_{\rm orb}
< 10$\,days.  We let $x = 1$ and 2.  We adopt $f(P_{\rm orb}) \propto
P_{\rm orb}^{-y}$ over 1--10\,days.  Multiplying the resulting value
of $\mathscr{E}$ by 1000 provides a crude estimate of the actual
number of {\em Kepler} targets with detectable ellipsoidal
variability.  No single value of $y$ is consistent with the data, and
so we consider the reasonable range $y = -1$, 0, and $+1$.
Inclinations are chosen under the assumption that the orbits are
randomly oriented, such that $f(\cos I) = 1/2$ for $I\in (0,\pi)$.
Our calculations use fixed values of $(\delta\fdisk/\fdisk)_{\rm int}
= 10^{-5}$ and $T_6 = n_{100} = 1$.

Results of our Monte Carlo integrations are shown in
Table~\ref{tab:detnum} as actual numbers of {\em Kepler} targets.  The
largest number of detectable systems is obtained when $x = y = 1$,
parameters that yield the largest proportions short periods and
massive planets.  We expect that $\sim$10--100 {\em Kepler} stars may
exhibit ellipsoidal oscillations with $S/N \ga 5$.  A handful of
systems might have $S/N \ga 10$.  Higher harmonics from the $\ell = 3$
components of eq.~(\ref{eq:fdiskeq}) or modest eccentricities might be
accessible for at most a few stars.

Our integrations also check for cases where the planet is transiting.
As $(S/N)_{\rm min}$ increases from 1 to 5, the fraction of systems in
Table~\ref{tab:detnum} with $|\cos I| < (R+R_p)/a$ runs from
$\simeq$30\% to $\simeq$50\%, with a weak dependence on $x$ and $y$.
Such significant fractions stand to reason, since systems with the
shortest periods have the highest ellipsoidal amplitudes and transit
probabilities.  Transit measurements directly give $P_{\rm orb}$,
$\sin I \ga 0.95$ (for $P_{\rm orb} \ga 1$\,day), and $(R_p/R)^2$.
The planet mass $M_p$ can be determined with the addition of
spectroscopic radial velocity measurements, which should be possible
for most of the {\em Kepler} targets with detectable ellipsoidal
oscillations.  The ellipsoidal amplitude then depends on the
unmeasured stellar mass and radius via $\pert \propto R^3/M^2$
(eq.~[\ref{eq:tidescale}]), as well as the stellar photospheric
conditions (eq.~[\ref{eq:simplesurfacedf}]).  If $M$ and $R$ are
obtained from stellar models, ellipsoidal variability may provide an
interesting consistency check on all the system parameters, as well as
test the theory of forced stellar oscillations.
 
As a last point, we emphasize that stars of mass $\ga$$1.4\msun$ may
have typical ellipsoidal amplitudes of $\sim$$10\pert$.  However, such
stars will also be younger than most {\em Kepler} targets and probably
have intrinsic variability $\gg$$10^{-5}$.  We carried out Monte Carlo
integrations with $M = 1.4$--$1.6\msun$, $(\delta\fdisk/\fdisk)_{\rm
  ell} = 10\pert\sin^2 I$, and $x = y = 1$.  As we vary
$(\delta\fdisk/\fdisk)_{\rm int}$ from $10^{-5}$ to $10^{-4}$,
$\mathscr{E}$ decreases from large values of $\simeq$0.4 to a small
fraction of $\simeq$0.03 for $(S/N)_{\rm min} = 10$.  Unfortunately,
we do not know how many such stars will be included in the {\em
  Kepler} target list.  Also, there has not yet been a discovery of a
giant planet with $P_{\rm orb} < 10$\,days around a star of mass
$\geq$$1.4\msun$, but exoplanet surveys tend to exclude these more
massive stars.

%--------------------------------------------------------------------------------

\acknowledgements 

We thank Tim Brown for general discussions and addressing {\em Kepler}
questions, J{\o}rgen Christensen-Dalsgaard for guidance on stellar
luminosity perturbations, and Mike Muno for advice on signal
processing.  This work was supported by NSF grant PHY05-51164.

%--------------------------------------------------------------------------------

\appendix

\section{OSCILLATION EQUATIONS}
\label{sec:equations}

Here we list the nonadiabatic, linearized fluid equations that we
solve numerically.  The reader is referred to \cite{Unno1989} for a
complete discussion.  Scalar and vector quantities are expanded in
spherical harmonics $Y_{\ell m}$ and poloidal vector harmonics,
respectively. The momentum, mass, and energy equations are written in
terms of the dimensionless variables $y_1=\xi_r/r$, $y_2=(\delta
p/\rho+\delta \varphi)/gr$, $y_3=\delta \varphi/gr$, $y_4=g^{-1}
d\delta \varphi/dr$, $y_5=\Delta s/C_p$, and $y_6=\Delta L/L$.  Here
$L$ is the total (radiative plus convective) luminosity. The radial
flux perturbation is $\Delta F/F=\Delta L/L-2\xi_r/r$.  In radiative
zones, the nonadiabatic equations are
\be
\frac{dy_1}{d\ln r}  & = &  y_1 \left( \frac{gr}{c_s^2}-3 \right)
+ y_2 \left( \frac{gk_h^2r}{\omega^2} - \frac{gr}{c_s^2} \right)
+ y_3 \frac{gr}{c_s^2} - y_5 \rho_s + \frac{k_h^2}{\omega^2} U ~,
\\
\frac{dy_2}{d\ln r}  & = &  y_1 \left(\frac{\omega^2-N^2}{g/r} \right)
+ y_2 \left( 1-\eta + \frac{N^2}{g/r} \right)
- y_3 \frac{N^2}{g/r} - \rho_s y_5 - \frac{1}{g} \frac{dU}{dr} ~,
\\
\frac{dy_3}{d \ln r} & = &  y_3 \left( 1 - \eta \right)
+ y_4 ~,
\\
\frac{dy_4}{d \ln r} & = & y_1 \eta \frac{N^2}{g/r}
+ y_2 \eta \frac{gr}{c_s^2}
+ y_3 \left[ \ell(\ell+1) - \eta \frac{gr}{c_s^2} \right]
- y_4 \eta + y_5 \rho_s \eta ~,
\\
\frac{dy_5}{d\ln r}  & = &  y_1 \frac{r}{H_p} 
\left[
\nabla_{\rm ad} \left( \eta - \frac{\omega^2}{g/r} \right)  
+ 4 \left( \nabla - \nabla_{\rm ad} \right) + c_2
\right]
+ y_2  \frac{r}{H_p} 
\left[
\left( \nabla_{\rm ad} - \nabla \right) \frac{gk_h^2r}{\omega^2} - c_2
\right]
\nonumber \\
&& + y_3 \frac{r}{H_p} c_2
+ y_4 \frac{r}{H_p} \nabla_{\rm ad}
+ y_5 \frac{r}{H_p} \nabla \left( 4 - \kappa_s \right)
- y_6 \frac{r}{H_p} \nabla
+ \frac{r}{H_p}
\left[
\nabla_{\rm ad} \left( \frac{dU/dr}{g} + \frac{k_h^2}{\omega^2} U \right)
- \nabla \frac{k_h^2}{\omega^2} U
\right] ~,
\label{eq:y5}
\\
\frac{dy_6}{d\ln r} & = &
y_1  \ell(\ell+1)\left(\frac{\nabla_{\rm ad}}{\nabla} - 1 \right)
- y_2 \ell(\ell+1)\frac{\nabla_{\rm ad}}{\nabla}
+ y_3  \ell(\ell+1)\frac{\nabla_{\rm ad}}{\nabla}
+ y_5 \left[
i\omega \frac{4\pi r^3 \rho C_p T}{L}  - \frac{\ell(\ell+1)}{\nabla}\frac{H_p}{r}
\right] ~,
\label{eq:y6}
\ee
where $c_s$ is the sound speed, $\eta = d\ln M_r/d\ln r$,
$c_2=(r/H_p)\nabla (\kappa_{\rm ad}-4\nabla_{\rm ad})+\nabla_{\rm
  ad}(d\ln \nabla_{\rm ad}/d\ln r + r/H_p)$, and we have ignored
energy generation terms.  Note that the tidal acceleration $-\grad U$
has been added to the momentum equations. In convection zones, we
ignore turbulent viscosity effects and replace the radiative diffusion
equation (eq.~[\ref{eq:y5}]) with the prescription $\Delta s =
\textrm{constant}$ (see \S~\ref{sec:conv}), or more precisely
\be
\frac{d}{dr} \left( y_5 C_p \right) =  0~.
\ee
Equation~(\ref{eq:y6}) still involves the total (convective plus
radiative) luminosity. We ignore energy generation and horizontal flux
perturbation terms, i.e. we ignore all terms with spherical harmonic
index $\ell$ in eq.~(\ref{eq:y6}) in convection zones.

At the center of the star, we require the solutions to be finite, and
also set $\Delta s=0$. At the surface, we set $\delta p=\rho g \xi_r$
and we require $\delta \varphi$ to decrease outward.  This boundary
condition is only approximate, as $g$-modes may propagate above the
convection zone for wave periods of $\ga$$4\, {\rm days}$ in our
$1M_\odot$ model. The final surface boundary condition is given by
eq.~(\ref{eq:dphotp}).  Care must be used in the radiative zone just
below the photosphere, since the entropy perturbation is far from the
quasi-adiabatic value.  If we solve the radiative diffusion equation
in this region, we find that the entropy increases by $\sim$10 orders
of magnitude in just a few grid points.  However, we regard this
behavior as unphysical, because the region at the top of the
convection zone is optically thin. To eliminate this unphysical
behavior, we set $\Delta s$ to a constant at such low optical depths.

%--------------------------------------------------------------------------------

\end{document}

%% file: tab1.tex
\begin{deluxetable}{p{-1mm}lcclcc}[b]
\tabletypesize{\footnotesize}
\tablecolumns{7}
\tablewidth{0pt}
\tablecaption{Limb Darkening Parameters
\label{tab:ldterms}}
\tablehead{
\colhead{} & \colhead{} & \multicolumn{2}{c}{$b_\ell$} & \colhead{} & \multicolumn{2}{c}{$c_\ell$} \\
\cline{3-4} \cline{6-7} \\
\colhead{$\ell$} & \colhead{} & \colhead{General $\gamma$} & \colhead{$\gamma = 3/5$} & &
\colhead{General $\gamma$} & \colhead{$\gamma = 3/5$} 
}
\startdata
2 & & $(1+\gamma)/[20(3-\gamma)]$ & 13/40 & & $3(1+3\gamma)[10(3-\gamma)]$ & 39/20 \\
3 & & $\gamma/[4(3-\gamma)]$ & 1/16 & & $3\gamma/(3-\gamma)$ & 3/4
\enddata

\end{deluxetable}

%% file: tab2.tex
\begin{deluxetable*}{p{2.3cm}cccc}
\tabletypesize{\footnotesize}
\tablecolumns{5}
\tablewidth{0pt}
\tablecaption{Periodic Flux Modulations
\label{tab:fluxmod}}
\tablehead{
\colhead{Variability} &
\colhead{} &
\colhead{Dominant} &
\colhead{Phase at} &
\colhead{} \\
\colhead{Source} &
\colhead{Amplitude\tablenotemark{a,b}} &
\colhead{Harmonic} &
\colhead{Maximum/Minimum\tablenotemark{c}} &
\colhead{References} 
}
\startdata
Ellipsoidal\tablenotemark{d}\dotfill & $2\times 10^{-5}m_p m P_1^{-2}\sin^2 I$ & $P_{\rm orb}/2$ & 0.25(0.75)/0.00(0.50)  & \nodata \\
Doppler\tablenotemark{e}\dotfill & $3\times 10^{-6}m_p m^{-2/3} P_1^{-1/2}\sin I$ & $P_{\rm orb}$ & 0.25/0.75 & 1  \\
Reflection\tablenotemark{f}\dotfill & $6\times 10^{-5} (\alpha/0.1) m^{-2/3} P_1^{-4/3}\sin I$ & $P_{\rm orb}$ & 0.50/0.00 & 2,3 \\
Transit \dotfill & $10^{-2} m^{-2}$ & $P_{\rm orb}$ & \nodata/0.00 & 4
\enddata

\tablerefs{
(1)~Loeb \& Gaudi 2003;
(2)~Seager, Whitney, \& Sasselov 2000; 
(3)~Sudarsky, Burrows, \& Pinto 2000;
(4)~Seager \& Mall{\'e}n-Ornelas 2003
}

\tablenotetext{a}{We assume that the orbit is circular in our estimates.}

\tablenotetext{b}{The dimensionless variables used are $m_p = M_p/(10^{-3}\msun)$, 
$m = M/M_\odot$ and $P_1 = P_{\rm orb}/1\,{\rm day}$.  We have assumed that the 
star and companion have respective radii of $R/R_\odot = m$ and $0.1\rsun$.}

\tablenotetext{c}{The phase is in the range 0--1, where at phase $0$ the planet is 
closest to the observer.} 

\tablenotetext{d}{Only the $\ell = 2$ component of eq.~(\ref{eq:fdiskeq}), with $\lambda_2 = 2$, is considered here.} 

\tablenotetext{e}{We approximate the amplitude as $4v_r/c$, where $v_r$ is the reflex 
speed of the star along the line of sight,  and the factor of 4 is approximately what one obtains 
for a $V$-band spectrum similar to the Sun.}

\tablenotetext{f}{Here $\alpha$ is the geometric albedo of the companion.  
The inclination dependence is an approximation for $I\simeq 90^\circ$ and the Lambert phase function.}

\end{deluxetable*}

%% file: tab3.tex
\begin{deluxetable}{clcclcclcc}
\tabletypesize{\footnotesize}
\tablecolumns{10}
\tablewidth{0pt}
\tablecaption{Number of {\em Kepler} Stars with Detectable Ellipsoidal Oscillations
\label{tab:detnum}}
\tablehead{
\colhead{} & \colhead{} & \multicolumn{2}{c}{$(S/N)_{\rm min} = 1$} & \colhead{} & \multicolumn{2}{c}{$(S/N)_{\rm min} = 3$} &  \colhead{} &  
\multicolumn{2}{c}{$(S/N)_{\rm min} = 5$} \\
\cline{3-4} \cline{6-7} \cline{9-10}\\
\colhead{$y$} & \colhead{} & \colhead{$x = 1$} & \colhead{$2$} & \colhead{} &
\colhead{$x = 1$} & \colhead{$2$} & \colhead{} & \colhead{$x = 1$} & \colhead{$2$}
}
\startdata
1 & & 240 & 166 & & 76 & 35 & & 33 & 13\\
0 & & 99 & 62 & & 26 & 12 & & 11 & 4 \\
-1 & & 33 & 19 & & 7 & 3 & & 2 & 1
\enddata

\end{deluxetable}